\documentclass[sigconf]{acmart}
\AtBeginDocument{%
  }

\usepackage{array}
\usepackage{pifont}
\usepackage{bm}
\newcolumntype{M}[1]{>{\raggedright\arraybackslash}m{#1}}
\usepackage{amsmath}
\usepackage{balance} 
\usepackage{booktabs}
\usepackage{subcaption}
\usepackage{multirow}
\usepackage{multicol}
\usepackage{makecell}
\usepackage{booktabs}
\usepackage{tabularx}
\usepackage{graphicx}
\usepackage{tablefootnote}
\usepackage{threeparttable}
\usepackage{siunitx}
\usepackage{dsfont}
\usepackage{enumitem}
\usepackage{lipsum}

\usepackage{xcolor}
\newcommand{\todo}[1]{{\color{red} [TODO: #1]}}     % more works
\definecolor{mypurple}{RGB}{128, 0, 128}
\newif\ifshowcategory
\showcategoryfalse % hide

\newcommand{\rebuttalpoint}[1]{%
  \ifshowcategory
    {\color{mypurple}[#1]}%
  \fi
}

\usepackage{soul}
\soulregister\cite7 % allow highlight with~\cite inside
\soulregister\ref7
\soulregister\item7
\soulregister\itemize7
\soulregister\description7
\soulregister\enumerate7
\soulregister\footnote7
\soulregister\todo7
\soulregister\underline7
\soulregister\bad7
\soulregister\rev7
\soulregister\fixed7

\newcommand{\toreview}[2][yellow]{%
  {\sethlcolor{#1}\hl{\leavevmode#2}}%
}
\definecolor{revision_color}{HTML}{ACE8FF}
\newcommand{\rev}[1]{#1}
\definecolor{fixed_color}{HTML}{FFFFFF}
\newcommand{\fixed}[1]{\toreview[fixed_color]{#1}}
\definecolor{bad_color}{HTML}{FF5555}
\newcommand{\bad}[1]{\toreview[bad_color]{#1}}

\copyrightyear{2026}
\acmYear{2026}
\setcopyright{cc}
\setcctype{by}
\acmConference[MM '26]{Proceedings of the 34th ACM International Conference on Multimedia}{November 10--14, 2026}{Rio de Janeiro, Brazil}
\acmBooktitle{Proceedings of the 34th ACM International Conference on Multimedia (MM '26), November 10--14, 2026, Rio de Janeiro, Brazil}
\acmDOI{10.1145/3767308.3836354}
\acmISBN{979-8-4007-2213-4/2026/11}

\begin{document}

% \title{Human-Aligned Face Understanding in Large Models via Perceptual and Psychological Cues}
% \title{FACET: Facial Attribute Contrastive Explanation for Human-Aligned and Steerable Perception}
% \title{SHAPE: Steerable Human-Aligned Face Perception with Contrastive Attribute Disentanglement}
%\title{AlignFace: Human-Aligned Face Perceptual Similarity via Disentangled Semantics and Steerable Representation}
% \title{AlignFace: Towards Human-Aligned Face Perceptual Metric via Semantic Concepts and Steerability}
\title{AlignFace: Human-Aligned Face Similarity Metric with Interpretable Concept Relations}

\begin{teaserfigure}
  \centering
  \vspace{-0.25cm}
  \includegraphics[width=.98\linewidth]{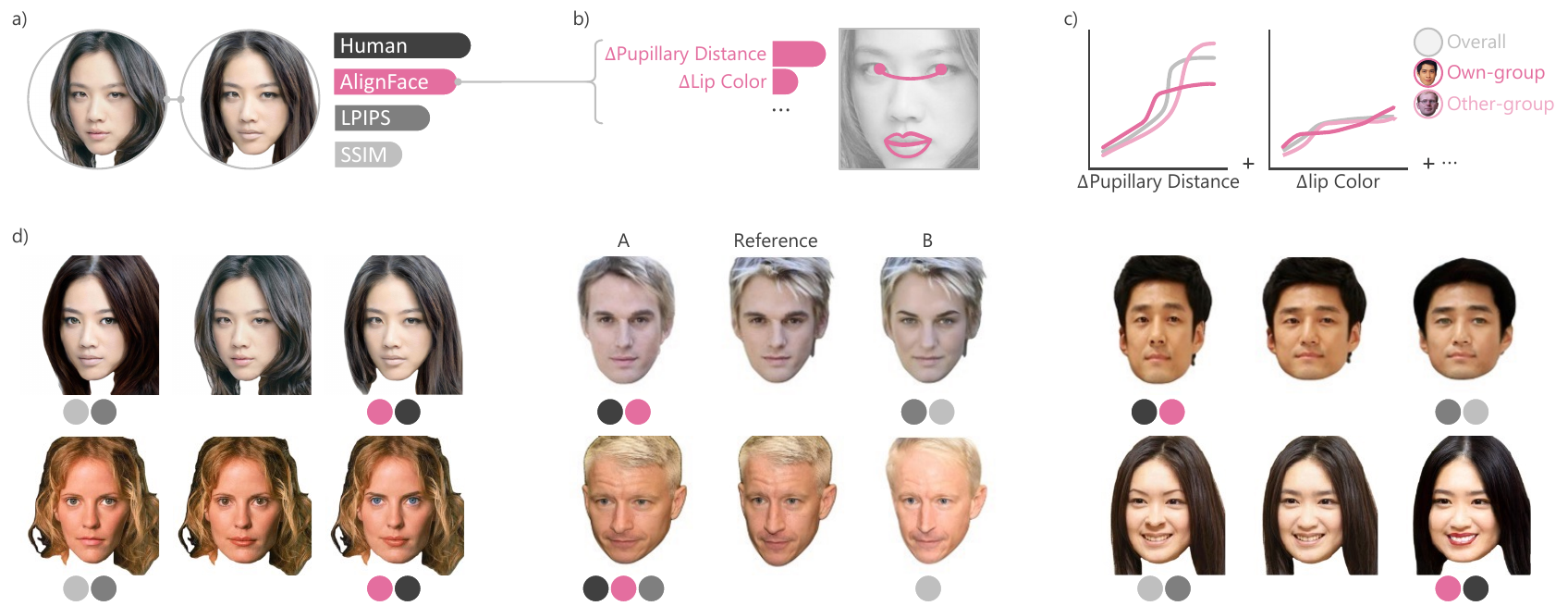}
  \vspace{-0.25cm}
  \caption{
    a) AlignFace provides a face-pair similarity score that better matches human perception. 
    b) It decomposes face similarity into attribute-level comparisons, capturing featural (e.g., lip color) and configural (e.g., pupillary distance) cues for fine-grained explanations. 
    c) Decisions are explained via learned nonlinear functions that map attribute similarity to overall face similarity, while capturing group-dependent effects (own- vs. other- group).
    d) AlignFace outperforms metrics like LPIPS and SSIM in triplet comparisons---whether A or B is more similar to a reference---improving alignment with human perception.
     % \todo{Update with 3x3, balance race, gender, age, with varing facial hair, glasses diff ... Also increase the font size ...}
     % Given the triplet task, ``Which face (A or B) is more similar to the reference?", existing metrics often failed to align with human ratings. 
     % Moreover, they lack the granularity to capture specific facial semantics (e.g., eye shape).
     % Our proposed AlignFace achieves superior alignment with human perceptual judgment across both face and attribute dimensions.
  }
  % \vspace{-0.1cm}
  \Description{A description for accessibility. This should describe what is shown in the figure.}
\end{teaserfigure}

% =========================
% Authors (ANONYMOUS REVIEW)
% =========================
% If double-blind: keep anonymous.
% Replace with real authors only for camera-ready.

% \author{Anonymous Author(s)}
% \affiliation{%
%   \institution{Anonymous Institution(s)}
%   \country{Anonymous Country}
% }
% \email{anonymous@anonymous.com}

% =========================
% Authors (CAMERA-READY)
% =========================
% Uncomment this block for camera-ready.
\author{Ying Huang}
\authornote{These authors contributed equally to this work.}
\affiliation{%
  \institution{Department of Computer Science, National University of Singapore}
  \city{Singapore}
  \country{Singapore}
}
\email{ying.h@u.nus.edu}

\author{Wencan Zhang}
\authornotemark[1]
\affiliation{%
  \institution{Department of Computer Science, National University of Singapore}
  \city{Singapore}
  \country{Singapore}
}
\email{wencanz@nus.edu.sg}

\author{Brian Y. Lim}
\affiliation{%
  \institution{Department of Computer Science, National University of Singapore}
  \city{Singapore}
  \country{Singapore}
}
\email{brianlim@nus.edu.sg}

% \renewcommand{\shortauthors}{Huang et al.}
% Optional short author list for headers (camera-ready only)
\renewcommand{\shortauthors}{Huang et al.}

\begin{abstract}
Computer vision models for generated facial content, such as face editing and privacy protection, increasingly affect people, requiring similarity metrics that serve as faithful proxies for human perception. While perceptual evaluation has progressed from signal-based heuristics to representation-based metrics, current approaches are limited to behavioral modeling without cognitive alignment. They rely on implicit and spurious relations while assuming a universal observer, failing to account for inherent variations across diverse human populations. This leads to inaccurate evaluative models of stakeholders and misleading guidance for generative model debugging. Rather than treating perception as a black box, we leverage scientific findings from cognitive psychology of human face similarity perception: dependence on facial featural and configural attributes, nonlinear psychophysical response scaling, and own-group biases. 
We introduce the FACETS dataset and propose AlignFace, an interpretable, human-aligned, face similarity metric that encodes these cognitive principles through ante-hoc modeling. 
It employs visual-language modeling (VLM) to encode paired face images and text-based attributes, gated cross-attention (CA) to extract attribute-specific facial difference representations, concept bottleneck modeling (CBM) to constrain reasoning via interpretable face attributes, and neural generalized additive model (GAM) to model their nonlinear influence. 
Experiments found AlignFace significantly improves alignment with human subpopulation perceptions compared to baseline metrics, including recent domain-free learned perceptual metrics. 
By bridging learned representations and human cognitive processes, this work enables more transparent and aligned perceptual evaluation metrics for face images.
\end{abstract}




\maketitle
% =========================
\section{Introduction}
The study of faces in computer vision has given rise to many applications for machine understanding, including
face recognition~\cite{schroff2015facenet, deng2019arcface, wang2018cosface, lin2021xcos}, social media retrieval~\cite{zaeemzadeh2021face, chong2021retrieve}, expression recognition for affective computing~\cite{wang2023rethinking, cui2025learning}, and descriptive captioning~\cite{jang2023thats, li2025faceinsight}.
In these paradigms, faces are treated as data points for classification. 
However, innovations in generative face synthesis and automated image editing have redefined the prediction tasks. 

Recent works target human observers with visual content for human consumption, instead of machine interpretation.
These human-centric applications range from face editing~\cite{lee2020maskgan, liang2021ssflow, shi2022semanticstylegan, huang2023collaborative, du2023pixelface+, bobkov2024devil}, including digital makeup~\cite{li2018beautygan, deng2021spatially, yang2022elegant} and restoration~\cite{wang2021towards, yue2024difface} to obfuscation for privacy protection~\cite{yuan2022pro, barattin2023attribute, yang2024once}.
% \bad{Evaluating model efficacy requires determining the similarity between images.}
While standard methods for evaluating model performance involve metrics and models~\cite{wang2004image, hore2010image}, these data-centric approaches are poor proxies for human perception~\cite{geirhos2018imagenet}, critical for estimating the true effects on people.
% \todo{Introduction: Inaccurate estimation of human proxy: may under-estimate the privacy risk, ID deviation/consistency(from GAN) with face editing; optimizing on the wrong features, misalign with human.}
%
% ---------- Rebuttal (wsLd), exact wording ----------:
% Motivation for human-proxy metric (wsLd). 
% Reply: "We clarify that AlignFace is meant
% to estimate human perception of faces, thus serving as a human proxy. It could be used to reveal that a high-performing face recognition/verification model deceptively underestimates privacy risk or identity shifts by identifying that 2 faces are different when a human would still perceive them as similar. Our primary aim is human-alignment rather than data-centric performance."
% \rebuttalpoint{Motivation for human-proxy metric (wsLd).}
% \toreview{An inaccurate estimation of human proxy may underestimate facial privacy risks or identity shift for face editing,
% leading to optimizing the wrong features and misaligning with human expectations.}
\rev{Thus, a high-performing face verification model could deceptively underestimate privacy risk or identity shifts by identifying two faces as different when a human would still perceive them as similar, or vice versa.}
Recent efforts to embed human perceptual ratings~\cite{zhang2018unreasonable}, semantic constraints~\cite{muttenthaler2023improving}, and conceptual abstraction~\cite{muttenthaler2025aligning} lack transparency and are not cognitively grounded, risking spurious correlations~\cite{ross2017right} and being unfaithful to human perception.
% \todo{emphasizing human proxies are necessary to evaluate those media content; current superhuman AI have biased estimate; we are improving human proxies}

Instead, we argue that perception models should adhere to scientific cognitive effects to guide their alignment with humans.
For face perception tasks, we focus on the psychology of human face recognition~\cite{bruce1986understanding}.
First, humans process faces based on featural attributes (face parts, e.g., eye color, nose shape) and configural attributes (spatial layout, e.g., pupillary distance, jaw width)~\cite{collishaw2000featural, maurer2002many, abudarham2016reverse}. 
This determines which features or concepts face perception models should encode explicitly or implicitly.
Second, humans do not perceive stimuli linearly~\cite{stevens1957psychophysical}, e.g., perceived brightness varies by the square root of light intensity, and perceived differences depend on absolute values of each stimulus~\cite{gescheider2013psychophysics}.
Hence, we hypothesize that human perception of attribute differences nonlinearly affects their perception of the similarity between two faces.
Third, humans have varied backgrounds and abilities, leading to individual or group variances across a population. 
Notably, own-group bias has been observed, showing that people better recognize faces from their own demographic groups (ethnicity, gender, etc.)~\cite{{meissner2001thirty, malpass1969recognition}}.
Thus, human alignment should steer model behaviors toward human subgroups rather than assume a universal model for all humans.

We leverage the aforementioned cognitive principles with a data collection experiment and a new human-aligned model for face perception.
We collected large-scale human perceptual annotations for both overall face and attribute-specific similarity ratings from two-alternative forced choice (2AFC) comparison tasks~\cite{bogacz2006physics}.

% ---------- Updated results after new participants ----------:
This produced \rev{9.36k} triplet ratings of face similarity from \rev{78} participants, and \rev{73.32k} triplet ratings of 20 face attributes from \rev{611} participants.
% across 480 face triplets.
Trained on these data and grounded in cognitive principles, we propose \textit{AlignFace}, a human-aligned similarity metric model for face perception.
It is ante-hoc interpretable with modules for face attribute predictions and nonlinear attribute-overall perception mapping.
Each face attribute is predicted from a contextual fusion of a pair of face images for comparison, and a textual prompt of the attribute (e.g., \textit{``eyebrow shape''})~\cite{antol2015vqa}.
The attributes are combined as a multi-label concept set in a Concept Bottleneck~\cite{koh2020concept} to ensure downstream decisions in terms of these interpretable attributes.
These attributes then serve as inputs into a neural Generalized Additive Model (GAM)~\cite{chang2021node} to capture their nonlinear partial influence on the final overall perception score.

% To evaluate the faithfulness and human-alignment of our approach, we conducted experiments comparing 
In experiments, we evaluated the human-alignment of 
AlignFace against heuristic similarity, learned perceptual, and cross-modal baselines.
We compared the correlation with human ground-truth ratings for overall and per-attribute similarity, for all humans and for specific ethnic groups, and performed an ablation study on the model modules.
Results showed that AlignFace was more correlated to human perception than baselines overall, per-attribute, and per-group.
Visual examination of partial dependence from the GAMs also revealed highly nonlinear relations, confirming the hypothesis of nonlinear attribute influence.

% \todo{Change the order of contributions to emphasize cognitively-grounding contribution but weaken dataset contribution}

% ---------- Rebuttal (r5Tm), exact wording ----------:
% Emphasize cognitive characteristics more clearly in contributions and discussion (r5Tm). 
% Reply: "Thank you, we have revised accordingly."
% \rebuttalpoint{Emphasize cognitive characteristics more clearly in contributions and discussion (r5Tm).}

% ---------- Rebuttal (r5Tm), exact wording ----------:
% Paradigm innovation (jvsS).
% Reply: "We clarify that our contribution is a
% \textit{novel human-aligned and interpretable metric} (\JeJP) with \textit{cognitive characteristics} (\rTm), providing \textit{principled backing} (\wsLd). This aims for architectural ante-hoc interpretability, rather than a single mechanism."

Our \textbf{contributions} are:
\begin{enumerate}[label=\arabic*), leftmargin=*, itemsep=0em, topsep=0.2em, parsep=0pt, partopsep=0pt]
    \item \rev{\textbf{Cognitive characteristics}---featural and configural attributes, nonlinear perceptual scaling, and demographic (White/Asian) own-group bias---identified for human face similarity perception.}
    % ---------- Paradigm innovation (\jvsS) ----------
    % Reviewer's comment: The core modules (VLM feature extraction, cross-attention, concept bottleneck, and GAM) are all mature, existing techniques with no original mechanism or paradigm.
    % Reply: "We clarify that our contribution is a \textit{novel human-aligned and interpretable metric} (\JeJP) with \textit{cognitive characteristics} (\rTm), providing \textit{principled backing} (\wsLd). This aims for architectural ante-hoc interpretability, rather than a single mechanism."
    \item \textbf{AlignFace}, a cognitively-grounded, interpretable ante-hoc model for face perception 
    % \toreview{that operationalizes these cognitive characteristics through architectural constraints instantiated by} 
    that integrates \rev{the cognitive characteristics via}
    Visual-Language encoding, Gated Cross-Attention, Concept Bottleneck and Neural GAM to enforce human alignment.

    % ---------- Rebuttal (jvsS), exact wording ----------:
    % Similarity metric for practical utility (jvsS). 
    % Reply: We clarify that AlignFace does predict pairwise difference \hat{d} = F(\Delta\hat{\bm{c}}) in Eq. 2 of Sec. 4.4 and can be used as a metric for that. Triplet agreement was used only for training and evaluating AlignFaceSim with human labels in the 2AFC task. We will clarify that AlignFace is the main similarity metric, and will rename AlignFaceSim to AlignFace2AFC."
    \item 
    % \rebuttalpoint{Similarity metric for practical utility (jvsS).} 
    \rev{\textbf{AlignFace2AFC}}, an explainable human-aligned face similarity metric that explains perception via the nonlinear contributions of cognitively-grounded attributes.
    \item \textbf{FACETS},
    % (Face Attributes for Comparative Evaluation with Triplet Similarity), 
    a human-annotated dataset of face triplets with similarity judgments at both overall and attribute levels across 20 cognitively-grounded face attributes.
\end{enumerate}

% =========================
\section{Related Work}
\label{sec:related_work}

% We discuss why current face models fail to capture true human perception, due to neglecting cognitive psychology effects, and how AI interpretability methods are promising to integrate them.
We examine why current face models fail to capture true human perception due to neglecting cognitive principles, and how AI interpretability methods can close this gap.

\textbf{Face Representation Models.}
Face models have been developed for myriad tasks, such as identity recognition~\cite{schroff2015facenet, deng2019arcface, wang2018cosface}, expression analysis~\cite{zhang2022learn}, landmark localization~\cite{xia2022sparse}, and attribute prediction~\cite{liu2015faceattributes}.
Many depend on face representation model backbones, particularly margin-based architectures like ArcFace~\cite{deng2019arcface} and CosFace~\cite{wang2018cosface}, or triplet-loss based models like FaceNet~\cite{schroff2015facenet}.
However, despite their reliance on human-annotated data, these models remain primarily data-centric, optimizing for categorical ground truths rather than accounting for the subjectivity and nonlinear response scaling inherent in perceptual attributes.
This limitation stems from face benchmarks relying on binary annotations that neglect the contrastive nuance and context dependence of human perception~\cite{huang2008labeled, liu2015faceattributes, terhorst2021maad}.
% \todo{more clearly explain how and why the proposed method advances beyond prior work}
% To address this, we incorporate cognitively-grounded attributes, collect relative similarity judgments via a 2AFC task, and employ an ante-hoc architecture to model these relations.
%
% ---------- Rebuttal (r5Tm), exact wording ----------:
% Explaining advances beyond prior work (r5Tm). 
% Reply: "We have revised Related Work, e.g., on Perceptual Similarity Metrics: Unlike prior metrics that are data-driven and not cognitively grounded, we leverage these cognitive principles to design a human-aligned face perceptual metric that explicitly structures facial similarity into facial attributes, with nonlinear psychophysical weighting and observer-dependent modulation, introducing cognitively grounded inductive biases."
% \rebuttalpoint{Explaining advances beyond prior work (r5Tm).}
\rev{In contrast, our model training collects human perception labels from pairwise 2AFC judgments of overall face perception and multiple face attributes, and measures viewer demographics to account for perception biases.}

\textbf{Perceptual Similarity Metrics.}
The advancement of image generation necessitates evaluation metrics faithful to human perception. 
Conventional metrics like PSNR~\cite{hore2010image} and SSIM~\cite{wang2004image} or transformer-based representations like CLIP~\cite{radford2021learning} and DINO~\cite{oquab2024dinov2} fail to capture semantics or fine-grained perceptual nuances, offering poor proxies for human judgment. 
% Yet, while l
Learned metrics like LPIPS~\cite{zhang2018unreasonable} or DreamSim~\cite{fu2023dreamsim} improve alignment, their black-box nature risks depending on spurious features and correlations~\cite{ross2017right}. 
Indeed, DNNs often exploit cues divergent from human perception~\cite{geirhos2018imagenet}. 
While recent work attempts to enforce representational alignment through local-constrained global transform~\cite{muttenthaler2023improving} or concept abstraction~\cite{muttenthaler2025aligning}, these methods are data-driven and not cognitively grounded.

Focusing on facial perception, humans 
prioritize a combination of featural (face parts) and configural (spatial relations) attributes~\cite{collishaw2000featural, maurer2002many},
and integrate them additively~\cite{anderson1968simple, treisman1980feature}.
Furthermore, their effects are governed by nonlinear psychophysical laws~\cite{stevens1957psychophysical, gescheider2013psychophysics} and modulated by observer-dependent variations, such as the own-group bias~\cite{meissner2001thirty, malpass1969recognition}, characterized by superior recognition of one’s own demographic group. 
% \todo{more clearly explain how and why the proposed method advances beyond prior work}We leverage these cognitive principles for a human-aligned face perceptual metric.
%
% ---------- Rebuttal (r5Tm), exact wording ----------:
% Explaining advances beyond prior work (r5Tm). 
% Reply: "We have revised Related Work, e.g., on Perceptual Similarity Metrics: Unlike prior metrics that are data-driven and not cognitively grounded, we leverage these cognitive principles to design a human-aligned face perceptual metric that explicitly structures facial similarity into facial attributes, with nonlinear psychophysical weighting and observer-dependent modulation, introducing cognitively grounded inductive biases."
% \rebuttalpoint{Explaining advances beyond prior work (r5Tm).}
\rev{Unlike prior metrics that are data-driven and not cognitively grounded, 
we leverage these cognitive principles to design a human-aligned face perceptual metric that explicitly structures facial similarity through facial attributes, with nonlinear psychophysical weighting and observer-dependent modulation, introducing cognitively grounded inductive biases.}

\textbf{Explainable Vision Models.}
Ante-hoc interpretability has been shown to improve model performance while supporting human interpretability~\cite{rudin2019stop}.
Unlike visual explainable AI (XAI) techniques, such as saliency maps~\cite{selvaraju2017grad, simonyan2013deep} that can be spurious~\cite{zhang2020interpretable, zhang2022debiased}, 
such approaches explicitly encode cognitive principles~\cite{zhang2022towards, abichandani2025robust} or domain knowledge~\cite{matsuyama2023iris, lim2025diagrammatization} to improve human alignment.
While concept-based explanations~\cite{koh2020concept, kim2018interpretability} can encode human-understandable concepts, they assume these concepts influence behavior linearly.
Instead, generalized additive models (GAM)~\cite{caruana2015intelligible, chang2021node} can explain the nonlinear influence of features on predictions.
In this work, we developed an ante-hoc interpretable model, leveraging concept bottleneck~\cite{koh2020concept} and GAM~\cite{chang2021node} together to align perceptual attributes nonlinearly to human perception.

Furthermore, much of explainable AI (XAI) has focused on discrimination tasks, with few methods specifically for similarity tasks.
% However, t
They focus on showing comparative saliency maps with similar regions~\cite{zhu2021visual}, saliency and similarity~\cite{lin2021xcos, zhao2021towards}, and landmarks of positive/negative contributions~\cite{eberle2020building}.
% xCos ~\cite{lin2021xcos} is closest to our work, covering explanation of faces and their similarity, but it is not cognitively-grounded.
Within the facial domain, research similarly focuses on saliency maps~\cite{yin2019towards, williford2020explainable}, and more recently on grouped concepts of face attributes~\cite{teotia2022interpreting} or inferred concepts~\cite{plesh2024discovering}.
% \todo{more clearly explain how and why the proposed method advances beyond prior work}
% Instead, we explain the nonlinear effects of cognitively-grounded attributes on similarity perception.
%
% ---------- Rebuttal (r5Tm), exact wording ----------:
% Explaining advances beyond prior work (r5Tm). 
% Reply: "We have revised Related Work, e.g., on Perceptual Similarity Metrics: Unlike prior metrics that are data-driven and not cognitively grounded, we leverage these cognitive principles to design a human-aligned face perceptual metric that explicitly structures facial similarity into facial attributes, with nonlinear psychophysical weighting and observer-dependent modulation, introducing cognitively grounded inductive biases."
% \rebuttalpoint{Explaining advances beyond prior work (r5Tm).}
% \toreview{However, these methods explain model evidence rather than perceptual mechanisms underlying human similarity judgments. We instead explain face similarity through cognitively grounded attributes and their nonlinear effects, showing how facial differences shape human perception.}
\rev{While prior methods focus on attribution of latent activations or features,
we ground and constrain our approach on cognitive principles, leading to a more theoretically justified model, which we also show to be more accurately aligned with human perception.}

% \todo{Missed multimodal methods, since wasn't clear why this was discussed in earlier draft. But now I recall that our technical approach extracts concepts from VLLM.}

% \subsection{Psychology Insights for Face Perception}
% % =========outline=========
% \outline{
% What inspiration can we obatin from psychology? \\
% - 19 insights of human face recognition for CV~\cite{sinha2007face} \\
% - Configural vs. featural processing matters \\
% - Own group bias triggers different face processing mechanisms:}
% % =========outline=========

% \subsection{Large Models for Face Understanding (Face-MLLM / Face+LLM)}

% - concepts: cognitive, aligned, perceptual, similarity, face, attibutes

% =========================
% \section{Dataset Curation}
% \section{Cognitively-Aligned Perceptual Similarity (CAPS) Dataset}

\section{Face Attributes for Comparative Evaluation with Triplet Similarity (FACETS) Dataset}
To develop a human-aligned face similarity metric, we collected 
human perception ratings of similarity between faces overall and per-attribute for 20 face attributes.
% a comprehensive face triplet dataset by eliciting human perceptual judgments across both overall face and face attributes.

% \subsection{Stimuli Preparation}
\subsection{Image Selection and Stimuli Preparation}
% % =========outline=========
% \outline{
% What inspiration can we obatin from psychology? \\
% - 19 insights of human face recognition for CV~\cite{sinha2007face} \\
% - Configural vs. featural processing matters \\
% - Own group bias triggers different face processing mechanisms:}
% % =========outline=========
% \begin{table}[!t]
% \centering
% \caption{Demographic distribution of face targets selected during dataset curation.}
% \renewcommand{\arraystretch}{0.8}
% \vspace{-0.3cm}
% \begin{tabular}{lcccc}
% \toprule
% \textbf{Race} & \multicolumn{2}{c}{Caucasian} & \multicolumn{2}{c}{Asian}   \\
% \midrule
% \textbf{Gender} & Female & Male & Female & Male \\
% \midrule
% \textbf{Multi-PIE} & 20 & 20 & 10 & 10 \\
% \textbf{CelebA} & 20 & 20 & 10 & 10 \\
% \bottomrule
% \end{tabular}
% % \vspace{-0.3cm}
% \label{table:dataset_distro}
% \end{table}

We constructed our face-pair stimuli using images from two face datasets spanning both lab-controlled and in-the-wild faces: CMU Multi-PIE~\cite{gross2010multi} and CelebA~\cite{liu2015faceattributes}.
% Face instances were selected to ensure a representative distribution while balancing across different demographic groups (gender and ethnicity), 
% ---------- 120 identities (jvsS) exact wording ----------
% "Improving on prior datasets on face psychology (e.g., \cite{abudarham2016reverse}), we aimed to construct a dataset with balanced ethnicities (White, Asian) to facilitate own/other-group evaluation. 
% However, the extremely small number of usable Asian identities (40) from the source datasets (MultiPIE~\cite{gross2010multi}, CelebA~\cite{liu2015faceattributes}) limited the total size, where we undersampled White faces for balance."
% resulting in 120 unique identities\footnote{\rebuttalpoint{Dataset with 120 identities (jvsS).} \toreview{Only 40 usable Asian identities were available in the source datasets, so we undersampled White identities to maintain a 2:1 White/Asian ratio to reduce ethnic imbalance.}} across two datasets (see Appendix Table~\ref{s_table:dataset_distro}). 
% To reduce potential confounding factors, we excluded images containing heavy facial hair or eyeglasses.
\rev{To better balance demographic groups (gender, ethnicity) from the two datasets, due to the severely limited number of Asian identities (only 40), we subsampled White identities (to 80) to obtain 120 identities (see Appendix Table~\ref{s_table:dataset_distro})}.

For each identity, we selected four source images with varying poses and illumination. 
This balances intra-identity variability with inter-identity variability across demographic groups.
% ---------- Attribute-validity (AC) exact wording ----------
% JeJP's comments: "Furthermore, the exclusion of faces with eyeglasses or heavy facial hair limits the metric’s ecological validity.”
% We didn't respond to this.
To reduce potential confounding factors, we excluded images containing heavy facial hair or eyeglasses.
% \footnote{\rebuttalpoint{Attribute-validity (AC)}\toreview{We excluded them for experimental control; future work should assess generalizability to these and other natural variations.}}.
This initial image set provides seed images that we use to generate edited faces for comparison.

\begin{table}[!t]
\centering
% \vspace{-0.2cm}
\begin{threeparttable}
\caption{
% face attributes used in the perception task, grouped into featural, configural, and surface \& external categories.
Cognitively-grounded featural and configural attributes of human face perception adapted\tnote{1} from~\cite{abudarham2016reverse}.
}
\label{tab:facial_attributes_no_region}
\small
\setlength{\tabcolsep}{5pt} 
\renewcommand{\arraystretch}{0.9}   
\vspace{-0.2cm}
\begin{tabular}{
    M{1.4cm}  % Category
    M{2.6cm}  % Attribute
    M{3.2cm}  % Scale
}

\toprule
\textbf{Category} & \textbf{Attribute} & \textbf{Scale (0--1)} \\
\midrule
\multirow{15.5}{*}{Featural}
% Featural
& Eyebrow shape     & Rounded--Straight \\
& Eyebrow thickness & Thin--Thick \\
\cmidrule(lr){2-3}
& Eye shape         & Narrow--Round \\
& Eye size          & Small--Large \\
& Eye color        & Light--Dark \\
\cmidrule(lr){2-3}
& Nose shape        & Pointed--Flat \\
& Nose size         & Small--Large \\
\cmidrule(lr){2-3}
& Mouth size        & Small--Large \\
& Lip thickness     & Thin--Thick \\
% & Lip color\tnote{1} & Light--Dark \\
& Lip color & Light--Dark \\
\cmidrule(lr){2-3}
& Hair color       & Light--Dark \\
% & Hair length       & Bald--Long \\
\cmidrule(lr){2-3}
& Skin color       & Light--Dark \\
& Skin texture      & Smooth--Rough \\
\midrule

\multirow{8.5}{*}{Configural}
% Configural
& Face aspect ratio & Short-wide--Tall-narrow \\
\cmidrule(lr){2-3}
& Hair length       & Bald--Long \\
\cmidrule(lr){2-3}
& Forehead height   & Short--Long \\
& Pupillary distance & Small--Large \\
\cmidrule(lr){2-3}
& Cheek shape       & Sunken--Puffy \\
& Jaw width         & Narrow--Wide \\
& Chin shape        & Pointed--Square\\
\bottomrule
% \multirow{4}{*}{\makecell[l]{Surface \& \\ External}}
\end{tabular}

\begin{tablenotes}
\setlength{\itemsep}{0pt}
\setlength{\parskip}{0pt}
\setlength{\parsep}{0pt}
\setlength{\topsep}{0pt}
\footnotesize
\item[1] We omitted ear attributes due to occlusion by hairstyles, and added lip color due to variation in makeup.
\end{tablenotes}

\label{tab:facial_attributes}
\vspace{-0.2cm}
\end{threeparttable}
\vspace{-0.2cm}
\end{table}
% \note{Justify that edited faces enable capturing the subtle differences among similar faces. Testing on the original faces can not.}

% Rather than comparing arbitrary faces\footnote{Too similar if from the same identity, or too unrelated if from different identity.}, 
To emulate real-world scenarios, where changes are often subtle to preserve identity (face editing or retouching) or contextual information (privacy protection),
we synthesized edited face images using face image generation techniques to produce variations of seed face photos.
We employed diffusion-based inpainting~\cite{podell2024sdxl} for editing featural attributes, landmark-based warping~\cite{lugaresi2019mediapipe} for configural attributes, and GAN-based transfer~\cite{nikolaev2024hairfastgan} for hair-related attributes while preserving hairstyle consistency with demographics. 
% Using \todo{attribute inference model}, 
Based on detected face landmark points, we estimated attribute values related to length, size, and shape using geometry-based heuristic face anthropometric methods~\cite{Farkas1994Anthropometry}.
Color-relevant attributes were estimated from image color histograms from cropped facial components.
We ensured that the generated face had the same distribution of attribute values as the original faces (see Appendix Fig.~\ref{s_fig:stats_edit_face}).
Finally, we further masked the background from the face images to avoid distraction bias during human annotation.
% This emulates real-world scenarios, such as face editing, privacy protection, where 
% users need to discern subtle modifications between highly similar versions of a single target.
% This emulates real-world scenarios where changes are often subtle to preserve identity (face editing) or contextual information (privacy protection).

% To simulate the subtle changes in face editing or restoration scenarios, we generated edited versions of the source face images using AI-powered editing tools. 
We constrained the edits to 20 face attributes identified as highly influential in face space~\cite{abudarham2016reverse}, encompassing both featural and configural attributes (Table~\ref{tab:facial_attributes}).
\textit{Featural} attributes describe discrete components or surface qualities that can be identified in isolation, such as the specific color of an eye or the smoothness of skin, regardless of their spatial position on the face.
\textit{Configural} attributes describe the spatial geometry and relational distances between features, ranging from internal gaps like pupillary distance to the external silhouette boundaries defined by attributes like cheek and chin shapes, and hair length.
Each source image underwent two distinct edits, each editing 3--15 face attributes with randomly sampled magnitudes. 

% We employed attribute-specific editing strategies, including diffusion-based inpainting for featural and skin attributes~\cite{podell2024sdxl}, landmark-based warping for configural attributes, and GAN-based transfer for hairstyle-related attributes~\cite{nikolaev2024hairfastgan}. 
% To preserve demographic consistency, donor faces for hair transfer were sampled from the same racial and gender groups as the targets. 
% We verified that the edited stimuli preserved the attribute distributions of the original datasets 
% We ensured that the generated face had the same distribution of attribute values as the original faces (see Appendix Fig.~\ref{s_fig:stats_edit_face}). \todo{Need to describe how attribute features are calculated based on landmark/geometry heuristic } 
% This validation step ensures that the edited faces remain realistic and avoids introducing unnatural or biased variations that could distort human perception or affect downstream model learning.
% Finally, we further cropped the background from the face images to mitigate potential response bias during human annotation. 

% \subsection{Face Perception Study}
% \label{sec:face_perception_study}
\subsection{Psychophysical Similarity Measurements} % same as LPIPS paper
\label{sec:user-study}

% =========outline=========
% \outline{
% Having measured ID, perceived face similarity is an important proxy metric to measure the quality of edited/retouched/restored face photo. (SSIM/PSNR/LPIPS) \\
% We carried out a comparative 2AFC~\cite{bogacz2006physics} test, A spatial 2AFC (side by side) \\
% Recruit people and how we obtain annotations.
% }
% =========outline=========

\begin{figure}[!t]
    \centering
    \vspace{-0.2cm}
    \includegraphics[width=0.95\linewidth]{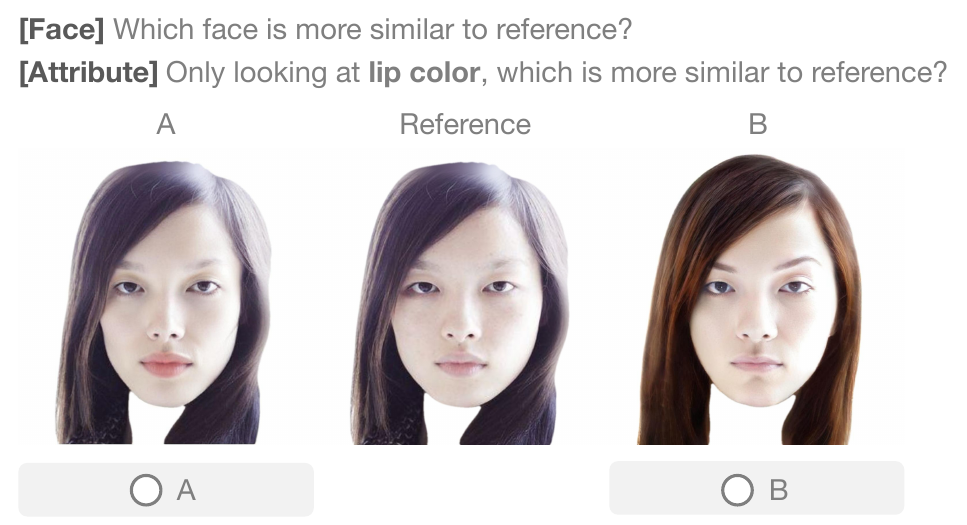}
    \vspace{-0.4cm}
    \caption{
    % The interface with 2AFC protocol in our two user studies.
    % For face perception task, participants chose the image more similar to a reference face. 
    % For the attribute task, they compared images based on a specified attribute (e.g., lip color).
    Experiment apparatus of the 2AFC protocol for two user studies on
    [Face] overall similarity and per-[Attribute] similarity, 
    with different survey questions based on the task.
    }
    \label{fig:exp_protocol}
    \vspace{-0.4cm}
\end{figure}
% \todo{Need to present triplet to show the protocol in our user study. Check how MM paper make their exp protocol figure more complex? (Ans: most MM papers don't have this figure. Some dataset curation paper report the statistics of the dataset.)}

% \note{
% Justification for 2AFC. It is widely deployed in perception study because:
% - can reduce response bias by leveraging this difference-based perception, more reliable comparisons than rating scales. \\
% - it converts an absolute judgment into a relative comparison. \\
% Because participants must choose between two options, it can avoid scale interpretation differences or subjective rating biases. \\
% - High sensitivity for small perceptual differences. \\
% How to construct face triplets? \\
% - (R) reference is the original unedited face. \\
% - (A, B) 2 options are two edited version from R. \\
% -- This is more ecologically valid for face applications with minor changes rather than comparing among difference targets and comparing difference face photos of the same target. }

% Having measured face identification, 
% Next, we investigated perceived face similarity to explore the perceptual gap between human observers and automated metrics.
Having prepared face image data, 
similar to data collection studies of human similarity perception~\cite{zhang2018unreasonable, fu2023dreamsim},
we elicited human perception ratings via
% To elicit these judgments, we conducted a 
the comparative two-alternative forced choice test (2AFC)~\cite{bogacz2006physics}. 
Instead of rating the similarity between two items (A and Target), the participant chooses which of two choices (A or B) is more similar to a Reference~\cite{parikh2011relative, thurstone2017law}.
This improves sensitivity to perceptual differences~\cite{so2023measuring} and mitigates response biases~\cite{wetzel2016response}.
% The 2AFC paradigm is widely used in psychological research because it relies on comparative rather than absolute judgments~\cite{parikh2011relative, thurstone2017law}.
% This approach effectively mitigates response biases~\cite{wetzel2016response} and offers higher sensitivity to subtle perceptual differences~\cite{so2023measuring}, resulting in more reliable comparison judgments.

% In each trial, given a reference face $R$, participants were presented with two candidate faces ($A$ or $B$) and asked to choose which one was more similar to $R$.
% ---------- Not from rebuttal ----------:
% To correct 120 x 2 x 2
% \rebuttalpoint{Demographic balance (rTm, wsLd, JeJP, jvsS).}
We constructed 480 
% \toreview{(120 identities $\times$ 4 source images)} 
face triplets 
\rev{(4 random tuples from each of 120 identities)}
% \todo{multiply is wrong, need to be corrected}
$\langle R, A, B\rangle$, where $R$ denotes the original unedited face, and $A$ and $B$ are two edited versions derived from $R$.
For each triplet, 
to assess overall similarity, we asked participants to choose \textit{``which face is more similar to the reference''};
to further assess per-attribute similarity, we asked participants to choose based on one randomly selected attribute out of 20.
% \footnote{This design is more ecologically valid for evaluating face-editing applications than comparisons involving unrelated individuals.}. 
% In total, we generated 480 face triplets. 
Fig.~\ref{fig:exp_protocol} shows the experiment apparatus, with different questions for two user studies on overall face similarity and per-attribute similarity.

% In each task, participants evaluated face similarity with the 2AFC question (Fig.~\ref{fig:exp_protocol}): \textit{``Which face ($A$ or $B$) is more similar to the reference ($R$)?"} 
% For both user studies, each participant completed 120 trials each of a randomly selected triplet over four sessions, with a 30-second break between sessions to reduce fatigue.

% ---------- Rebuttal (wsLd), exact wording ----------:
% Motivation for human-proxy metric (wsLd). 
% Reply: "We have validated our original findings by recruiting additional Asian participants to mitigate demographic imbalance, improving the ethnic White:Asian ratio from 361:22 ($\sim$ 15:1) to 361:191 ($\sim$ 2:1). Perfect balance remains challenging due to the inherent ethnic bias in Western crowdsourcing platforms (e.g., Prolific, MTurk)~\cite{douglas2023data}. The expanded dataset yields results consistent (not significantly different) with those obtained from the original White-dominant dataset across face perception (e.g., AF-FLIP original vs. expanded: 89.5\% vs. 87.0\%), attribute perception (75.0\% vs. 79.4\%), and attribute influence alignment and subpopulation alignment (Section~\ref{sec:main_evaluation})."
% \rebuttalpoint{Demographic balance (rTm, wsLd, JeJP, jvsS).}
For both user studies, the experiment procedure is:
introduction, consent, screening questions to verify correctness on trivially easy cases, main study with 120 trials each of randomly selected triplets over four sessions interspersed with 30-second breaks, conclusion with demographic questions.
Each main study session included one attention check question similar to the screening question.
We recruited online participants from Prolific.
% ---------- Rebuttal (r5Tm, wsLd), exact wording ----------:
% "Nevertheless, with variation (120 IDs $\times$ 2 edits $\times$ 2 choices), our dataset contains a large 480 triplet instances, which is \textbf{sufficient for the learning task (\rTm)}---finetuning pretrained VLMs on faces, achieving high agreement (87.0\%). Consequently, \textbf{only 41 participants (\wsLd)} was sufficient for the overall face similarity to label the triplets, producing 5,400 ratings."
% \rebuttalpoint{sufficient for the learning task (r5Tm) \& only 41 participants (wsLd).}
\rev{78} participants\footnote{\rev{Although modest, this participant sample yielded 8,880 ratings across 480 triplets, sufficient to fine-tune the pretrained VLM and achieve high agreement (87.0\%) with human judgments. (see Fig.~\ref{fig:results_alignment_face}).}} 
completed the face similarity study in a median time of \rev{23.2} minutes and were compensated \pounds4.00, and 
\rev{611} participants completed the attribute similarity study in a median time of \rev{30.5} minutes and were compensated \pounds4.00.
Notable for our own-group analysis, we report the ethnicity distribution of our participants\footnote{Achieving perfect balance remains challenging due to the inherent demographic skew of Western crowdsourcing platforms~\cite{douglas2023data}.}: 
% with 227 males and 248 females (5 non-binary), with ages ranging from 21 to 81 (Median = 46.0).
% \todo{are you sure own-group is age-based and not ethnicity?}
\rev{377 White, 222 Asian, 55 Black, 11 Hispanic, and 24 Mixed-race.}
See Appendix Table~\ref{s_table:participants} for participant details.
The user studies were approved by our institutional review board.
% \todo{Update after the new recruitment.}

After collecting the human similarity ratings, 
to ensure data quality, we excluded \rev{6,510} trials with few ratings (fewer than 3), 
% \todo{Refer back to LPIPS, DreamSim to justify why exclude those; But also mention Juri learning, PSE to show our knowledge.}
had ambiguous ratings (M = 30--70\% selecting either A/B), or were in the same session as a failed attention check question (all 30 responses excluded).
Hence, from both user studies, we collected \rev{8,880} triplet ratings of face similarity, and \rev{72,450} triplet ratings of 20 face attributes.
We randomly selected 80\% of this dataset for model training, and 20\% for testing.

\subsection{Cognitive Characteristics of Face Perception}
\label{sec:empirical_cognitive}

\begin{figure}[!t]
    \centering
    \vspace{-0.2cm}
    \includegraphics[width=1.0\linewidth]{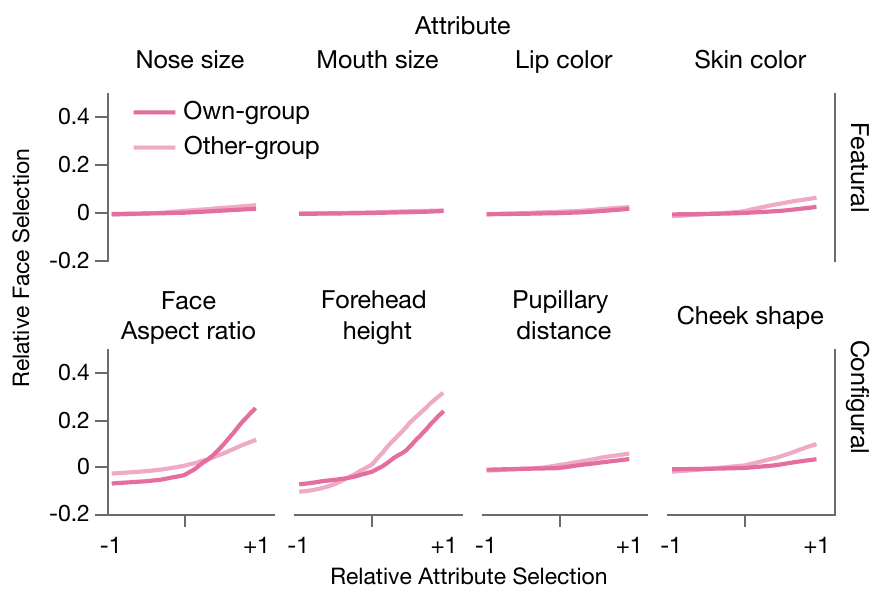}
    \vspace{-0.4cm}
    \caption{
    % \todo{Group space vertical and horizontal, with +1 and -1}
    Partial dependence plots of overall relative face selection by the 8 most salient attributes, split by own- and other-group.
    % , comparing human GAMs fitted on own-group and other group human judgments. 
    See Appendix Fig.~\ref{s_fig:gam_alignment_full} for full attributes.
    }
    \vspace{-0.4cm}
    \label{fig:human_GAMs}
\end{figure}
\begin{figure*}[!ht]
    \centering
    \vspace{-0.2cm}
    \includegraphics[width=1.0\linewidth]{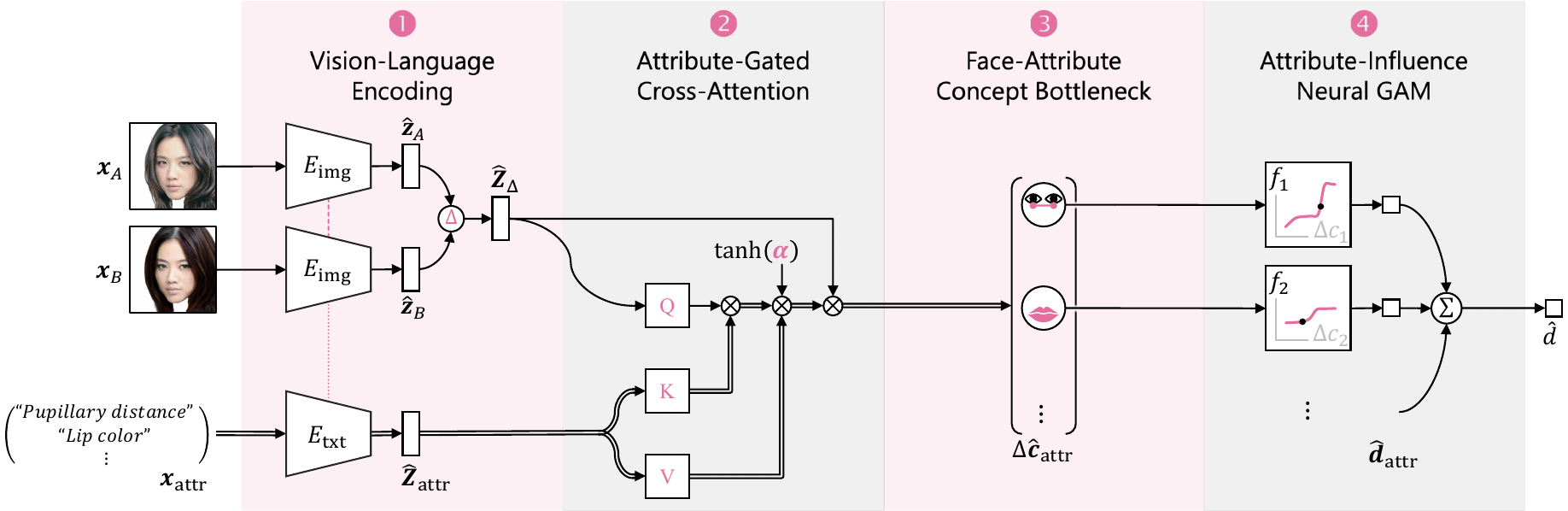}
    \vspace{-0.3cm}
    \caption{
        AlignFace architecture.
        1) vision-language model (VLM) encoding of face-pair images and text-based attributes into a shared representation,
        2) attribute-gated cross-attention (CA) to extract attribute-specific visual features,
        3) face-attribute concept bottleneck model (CBM) to constrain reasoning through disentangled face attributes, and
        4) Attribute-Influence neural Generalized Additive Model (GAM) to represent each attribute's nonlinear, additive contribution to the pairwise distance $\hat{d}$.    
        % \todo{rewrite caption} AlignFace architecture comprising: 
        % A visual encoder represents a face triplet as $\langle x_R, x_A, x_B \rangle$, while a textual encoder represents each attribute as $z^i_{T}$.
        % Embeddings are fused to predict attribute-conditioned distances $d_{A, R}^{i}$ and $d_{B, R}^{i}$; 
        % their difference yields the attribute preference $\Delta d_{p}$.
        % A concept bottleneck model aggregates $\Delta \boldsymbol{d}_{p}$ into overall face preference $\Delta d_{w}$ using GAM. 
    }
    \vspace{-0.3cm}
    \label{fig:architecture}
\end{figure*}

% % To assess attribute-level perceptual influence, 
% % To examine how attribute differences influence face similarity perception,
% To understand which attributes influences face similarity perception more than others,
% we fit a linear mixed-effects model (LMER) with 
% % human attribute 2AFC responses as the dependent variable. 
% overall 2AFC rating as response,
% % race-gender own-group, gender own-group, and attributes as fixed effects along with their interactions, 
% attribute type and ethnicity own-group as fixed effects along with their interactions, 
% and participants and triplets as random effects. 
% See Fig.~\ref{s_fig:human_group_bias} for results.

% However, we hypothesize that attribute differences nonlinearly affect overall perception,
% hence, we further fit
% % To examine the underlying cognitive principles of face perception at the reasoning level, we further analyzed aggregated ground-truth responses to characterize the effects of attribute relevance, nonlinear scaling, and demographic variance. 
% % In particular, we modeled human ratings of overall and per-attribute similarity using 
To understand which attributes influence face similarity perception more than others, we fit
Generalized Additive Models (GAM) with a logit link function. %with 
2AFC selection (A or B) as response and difference in attribute change (R$\rightarrow$A $-$ R$\rightarrow$B) as factors. 

% AC original comment: (b) scoping the own-group claims to the White/Asian comparison and hedging them to match the still-modest sample;
% \rebuttalpoint{Scoping own-group claims to White/Asian comparison (AC).} 
\rev{Given the demographic imbalance in our participant sample, we restrict the ethnicity-based own/other-group analysis to White and Asian participants.}
We fit two GAMs for own-group and other-group demographics to account for subpopulation variance. See Fig.~\ref{fig:human_GAMs} for the results.
% by comparing the ethnicity of viewers and stimuli (See Fig.~\ref{fig:human_GAMs}). This formulation enables us to examine nonlinear influence of the relevant attributes across human subpopulations.
% We discuss key insights next.

\textbf{Attribute Relevance.} 
Human similarity judgments were unevenly distributed across face attributes; certain features (e.g., face aspect ratio, forehead height) are significantly more influential than others (e.g., mouth size, lip color).
% This motivates a learnable weighting mechanism to prioritize highly informative attributes. \par

\textbf{Additive Nonlinear Scaling of Attribute Differences.} 
Perceptual sensitivity to attribute differences is strongly nonlinear.
Interestingly, as Relative Attribute Distance Difference increases, i.e., Face B is more different from the Reference than A, Relative Face Selection (toward B) increasingly increases.
However, this trend is reversed toward Face A, where the Relative Face Selection effect diminishes with Relative Attribute Distance Difference.
This suggests a side-choice bias~\cite{yeshurun2008bias}.
% \todo{Not only describe the trends of GAM, but say it is a good thing.}
% \todo{For example, sensitivity to “face aspect ratio” remains low until a threshold is reached, after which it increases sharply; “cheek shape” exhibits a sigmoidal (S-shaped) response pattern. } \par

\textbf{Demographic Variance (Own-Group Effect).} 
% The LMER reveals a significant interaction effect between perceived attribute difference and own/other-group (p $<$ .0001).
% , indicating that perceived attribute differences vary significantly depending on whether faces belong to the viewer’s own-group or other-group.
% For identical triplets, own-group viewers prioritize internal features---particularly eye-related attributes---as more defining of similarity compared to other-group viewers (p $=$ .0008), whereas this trend often reverses for nose-related attributes (p $=$ .0008). 
% Consistently, partial dependence plots differ behaviors for own-group or other-group GAMs.
% \rebuttalpoint{Scoping own-group claims to White/Asian comparison (AC).} 
\rev{For the White and Asian ethnic groups,} different trends were found for own-group and other-group ratings.
For example, humans were more sensitive to configural attributes Face aspect ratio and Forehead height, when perceiving faces sharing their own demographics (own-group).
Conversely, they were more sensitive to featural attributes Skin color and Cheek shape when perceiving other-group faces.
This perceptual divergence confirms that a single, universal similarity metric is insufficient for human-aligned evaluation, motivating the need for demographic-aware steering.

These findings of cognitive effects directly motivate our technical approach, which we describe next.

% \section{Technical Approach}
% \section{AlignFace: Human-Aligned Perceptual Metric}
% \section{AlignFace: Human-Aligned Perceptual Face Similarity Metric}
\section{AlignFace for Perceptual Similarity Metric}
\label{sec:alignface_model}
% \section{AlignFace Interpretable Perceptual Similarity Metric}
% \section{AlignFace: Interpretable Human-Aligned Perceptual Face Similarity Metric}
% \section{AlignFace Interpretable Perceptual Similarity}

% \todo{the current way to frame \textbf{Concept Bottleneck} model (CBM)~\cite{koh2020concept} can be improved.\\
% Update the writing after updating architecture figure.}

We propose an interpretable, human-aligned face similarity metric model---AlignFace---for both overall and attribute perceptual comparisons between faces (Fig.~\ref{fig:architecture}). 
% Grounded on cognitive psychology of human face perception, our framework first leverages vision-language models (VLMs)~\cite{radford2021learning, li2024flip} to encode attribute-specific textual semantics.
Grounded in cognitive principles identified in Section~\ref{sec:related_work} and observed in Section~\ref{sec:empirical_cognitive}, AlignFace predicts similarity in a face pair, accounting for multiple face attributes, their nonlinear scaling effects, and demographic-specific variance.

AlignFace leverages
1) vision-language models (VLMs)~\cite{radford2021learning, li2024flip} to map face images and extensible, open-ended attribute-based text prompts into a shared semantic embedding space,
2) gated cross-attention~\cite{alayrac2022flamingo} to extract attribute-specific visual features,
3) concept bottleneck model (CBM)~\cite{koh2020concept} to constrain reasoning through disentangled face attributes,
and 4) neural Generalized Additive Model (GAM)~\cite{chang2021node} to resolve overall similarity into an additive composition of nonlinear functions over individual attribute differences.
% To enhance interpretability, we employ a concept bottleneck model (CBM)~\cite{koh2020concept} to explicitly map face attributes onto intermediate concepts, enabling granular comparisons across diverse attributes.
% These representations are further calibrated via semantic metric learning on human-annotated triplets.
% Finally, we incorporate a neural Generalized Additive Model (GAM)~\cite{chang2021node} to bridge the nonlinear relationship between attributes and overall similarity, yielding a reliable yet inherently interpretable face similarity metric.
\rev{See Appendix Table~\ref{tab:design_rationale} for justifications of each module in AlignFace compared to standard alternatives.}

\subsection{Vision-Language Model (VLM) Encoding}
% =========outline=========
% \outline{
% - Need text encoders of CLIP/FLIP~\cite{radford2021learning, li2024flip} to guide the direction of attribute embeddings. \\
% -- Visual-encoder: obtain the face-pair embeddings $z_{RA}$ or $z_{RB}$. \\
% -- Textual-encoder: \textbf{Disentanglement via Semantic Prompting} For each attribute in the set of face attributes, we prompt the textual-encoder with paired prompts with opposite direction to obtain $z^i_p$ or $z^i_n$ (positive or negative anchors). \\
% -- At training time: we assign a 2AFC task for the model to learn if R is closer to A or B, where we use human annotations on the triplet to supervise. \\
% -- At inference time: the face-pair embedding and an attribute are input as the query. the face-pair is compared with reference node (e.g., "Same eye color" vs. "Different eye color") to determine the similarity.
% }
% =========outline=========
\begin{figure}[!ht]
    \centering
    % \vspace{-0.2cm}
    \includegraphics[width=0.95\linewidth]{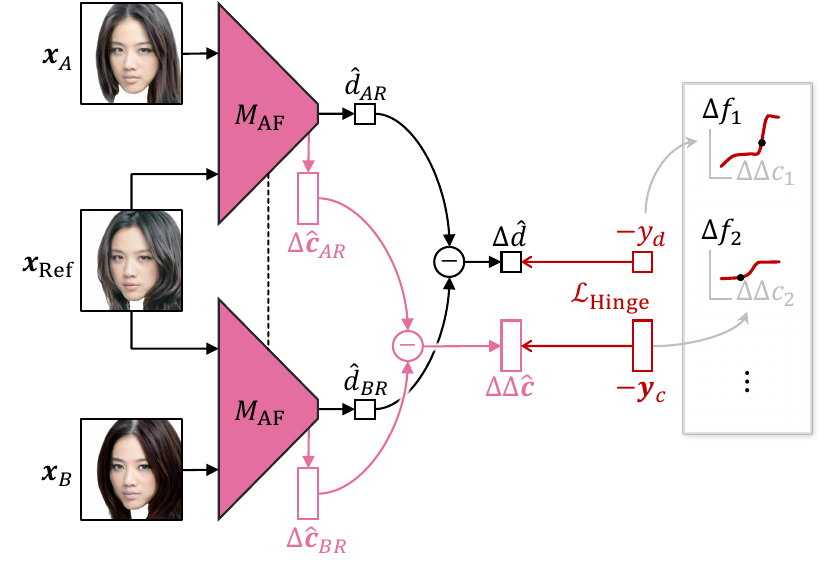}
    \vspace{-0.2cm}
    \caption{
        % \todo{rewrite caption}
        \rev{AlignFace2AFC} metric using AlignFace ($M_\text{AF}$) with a triplet instance $\langle x_A, x_\text{Ref}, x_B \rangle$ to predict two sets of overall and attribute distances and compute their differences ($\Delta \hat{d}$, $\Delta\Delta \hat{\bm{c}}$).
        Differences are learned via supervised training with binary 2AFC human labels of the overall face $y_d$ or of all attributes $\bm{y}_c$.
        See $y_d$--$\bm{y}_c$ GAM-based relations in Fig.~\ref{fig:human_GAMs}.
    }
    \vspace{-0.2cm}
    \label{fig:architecture_triplet}
\end{figure}
While a standard visual encoder extracts embeddings that capture overall facial information, these representations are often semantic-agnostic and thus fail to support attribute-targeted comparisons. 
We instead leverage a shared semantic representation space from a pre-trained VLM to enable face comparison along specific attribute directions (Fig.~\ref{fig:architecture}.1).
% ---------- Rebuttal (e31B), exact wording from Table R.1 ----------:
% "Vision-Language Model vs. Feature Engineering: More scalable: support open-domain (label-free) attributes instead of hard-coded features."
% \rebuttalpoint{Justify specific designs of each module (e31B).}
% \toreview{Unlike featural engineering, VLMs enable scalable and open-domain (label-free) attribute extraction through language prompts rather than hard-coded features.} 
% ---------- Rebuttal (e31B), exact wording from Table R.1 ----------:
% "We validated that general-purpose VLMs encode facial configural attribute information by comparing AlignFace's predicted attribute distances against ground-truth heuristic inter-landmark distances~\cite{lugaresi2019mediapipe}. Table~R.\ref{tab:manipulation_check} shows high correlation performance." 
% \footnote{\rebuttalpoint{Experiment to verify attribute detection (wsLd, jvsS)}\toreview{We further validated the VLM's ability to encode facial attributes using single-attribute manipulations (Appendix Table~\ref{tab:manipulation_check}).}}.
Specifically, we employ a shared vision encoder to extract representations for a face pair as $\langle \hat{\bm{z}}_A, \hat{\bm{z}}_B \rangle$.
We then apply relational reasoning~\cite{santoro2017simple} to model the differences between two face images, yielding $\hat{\bm{Z}}_{\Delta}$\footnote{Concatenation $A \oplus R$, element-wise difference $A-R$, and Hadamard product $A \odot R$.}. 
Next, we use the corresponding text encoder to extract descriptions of face attributes (e.g., \textit{``Pupillary distance''}) into $\hat{\bm{Z}}_{attr}$, which serves as an anchor guiding the comparison between the two faces along a specific attribute dimension.

% During training, we formulate a two-alternative forced choice (2AFC) task, where the model learns to predict whether a reference face $R$ is more similar to face $A$ or $B$, supervised by human annotations on triplet comparisons. At inference time, the model takes a face-pair embedding together with a queried attribute, and determines similarity by comparing the embedding against attribute-specific reference nodes (e.g., "same eye color" versus "different eye color"), enabling attribute-level perceptual comparison.

% \subsection{Concept Bottleneck Model via Attribute-Gated Cross-Attention}
\subsection{Attribute-Gated Cross-Attention (CA)}
% =========outline=========
% \outline{
% - \textbf{Semantic Alignment:} a game of attract and repel. \\
% -- Face pair embeddings: $z_{A, R|T_i}$ and $z_{B, R|T_i}$ \\
% -- Extract the face pair's conditional similarity: $s(A, R|T_i)$ and $s(B, R|T_i)$
% -- We learn the model's relative similarity on attribute $i$ as $\delta_i^A=s(A, R|T_i) - s(B, R|T_i)$ \\
% -- where $\delta_i^A>0$ means $A$ is closer to $R$ than $B$ in terms of attribute $i$ \\ 
% - We supervise this option score $\mathcal{L}^i(\tilde{s}^i, \mathcal{A}^i)$ with human response on each attribute $i$ using hinge loss, treated as a multi-task $\mathcal{L}=\Sigma^N_{i=1} \mathcal{L}^i$. \\
% -- A triplet $(x_R, x_A, x_B)$ with the target attribute $i$ to compare, the model's vote with a modified hinge loss~\cite{fu2023dreamsim}:\\
% \begin{equation}
% \mathcal{L}^i(\tilde{s}^i, \mathcal{\tilde{A}}^i) = max(0, m-\tilde{s}^i \cdot  \mathcal{\tilde{A}}^i),
% \end{equation}
% }
% =========outline=========

Having obtained the visual and textual embeddings, we perform multi-modal fusion to extract attribute-specific information for face comparison. 
In particular, we employ Attribute-Gated Cross-Attention (AGCA)~\cite{alayrac2022flamingo}
\rev{on image-difference features as queries (Q) and textual features as keys (K) and values (V), like in latent diffusion models~\cite{rombach2022high}\footnote{\rev{This inverted from cross-attention in text-guided VLMs~\cite{alayrac2022flamingo}, since we require attention in terms of face attributes (V) for downstream concept-bottleneck modeling. 
The image-diff query (Q) asks which attributes (K) explain the differences between the faces.}}}, 
%
% ---------- Rebuttal (jvsS), exact wording ----------:
% QKV Key/value attention design (jvsS). 
% "Using image-difference features as queries and text as keys/values is well-established (e.g., latent diffusion model (LDM)). AlignFace adopts a similar design, where we require attention outputs to represent face attributes (V) for the downstream concept-bottleneck modeling. The image-difference query (Q) asks which attributes (K) explain the difference for downstream concept prediction. This differs from cross-attention in text-guided VLMs."
% \rebuttalpoint{QKV Key/value attention design (jvsS).}
% \toreview{using the image-difference features as queries and textual features as keys and values, following the design of latent diffusion models~\cite{rombach2022high}}
(Fig.~\ref{fig:architecture}.2):
\begin{equation}
\hat{\bm{Z}}'_{\Delta}=\hat{\bm{Z}}_{\Delta}+\tanh(\alpha) \cdot \texttt{CrossAttn}(Q=\hat{\bm{Z}}_{\Delta}, K=\hat{\bm{Z}}_\text{attr}, V=\hat{\bm{Z}}_\text{attr}),
\label{eq:attn}
\end{equation}
where $\alpha$ is a learnable parameter for gating factor $\tanh(\alpha)$, controlling the attribute-conditioned signal injected into $\hat{\bm{Z}}_{\Delta}$.

% ---------- Rebuttal (e31B), exact wording from Table R.1 ----------:
% "Attribute-Gated Cross-Attention vs. Feature Concatenation: Interpretability and dimensionality reduction: constrain reasoning via attention to focus on most relevant information relevant to each attribute, rather finding relationships across the full face pair without differencing or focusing."
% \rebuttalpoint{Justify specific designs of each module (e31B).}
% \toreview{Rather than naive feature concatenation,}
The cross-attention operates analogously to a dictionary lookup, retrieving attribute-relevant visual features from $\hat{\bm{Z}}_{\Delta}$ conditioned on the attribute context $\hat{\bm{Z}}_\text{attr}$. 
% \toreview{The learnable gate adaptively constrains reasoning via attention to focus on most relevant information tribute, rather finding relationships across the full face pair without differencing or focusing.}
The residual connection stabilizes optimization by preserving the original representation.

\subsection{Face-Attribute Concept Bottleneck Model}
For a given face pair $\langle x_A, x_B \rangle$, the attribute-gated CA produces one attribute distance measure per text-based attribute prompt.
% module produces disentangled, attribute-specific representations by retrieving various face attributes. 
We denote these as \textit{interpretable} multi-label binary concepts $\Delta \bm{\hat{c}}_\text{attr} \in [-\bm{1}, +\bm{1}]$, which also bottlenecks subsequent reasoning (Fig.~\ref{fig:architecture}.3), making the architecture a concept bottleneck model (CBM)~\cite{koh2020concept}.

% ---------- Rebuttal (e31B), exact wording from Table R.1 ----------:
% "Concept Bottleneck Model vs. MLP: Interpretability: regularizes reasoning with domain-specific attributes, avoiding spurious features."
% \rebuttalpoint{Justify specific designs of each module (e31B).}
% \toreview{This bottleneck regularizes face similarity reasoning with domain-specific attributes, avoiding spurious features.}
However, unlike typical CBMs that input the concepts into an MLP for downstream reasoning, we leverage an interpretable module, described next.
% These representations are subsequently mapped to the concept bottleneck layer with an MLP (Fig.~\ref{fig:architecture}.3), where each attribute difference $\Delta \bm{\hat{c}}_{attr}$ can be used to guide and constrain the reasoning process.

\subsection{Attribute-Influence Neural GAM}
\label{sec:GAM_module}
To interpret the attribute influence while maintaining cognitive-grounding with nonlinear response scaling and additive attribute integration, 
we represent the relationship between overall distance and attribute distances with a Generalized Additive Model (GAM):
% ---------- Rebuttal (e31B), exact wording from Table R.1 ----------:
% "Neural GAM vs. MLP: Interpretability: regularizes reasoning independently and nonlinearly for each attribute, avoiding spurious relations."
% \rebuttalpoint{Justify specific designs of each module (e31B).}
% \toreview{we adopt a GAM to regularize reasoning independently~\cite{ashby1986varieties} and nonlinearly~\cite{yeshurun2008bias}:}
% for each attribute, thereby avoiding spurious relations:}
% We implement GAM within a neural network framework to preserve additivity across attributes while allowing for nonlinear transformations of individual features (Fig.~\ref{fig:architecture}.4):
\begin{equation}
F(\Delta \bm{\hat{c}}) = \sigma(\beta + \sum_{i=1}^{N} f_i(\Delta \hat{c}^i)),
\label{eq:ngam}
\end{equation}
where each $f_i$ is a spline function implemented via a small neural network, capturing both value-dependent attribute contributions and their relative importance across factors.
$\sigma(\cdot)$ denotes a sigmoid function\footnote{An inverse of the logistic link to model the bimodal distribution of human responses.},
and $\beta$ is the learned bias term.
To ensure perceptual consistency, we further impose a monotonicity constraint, i.e., $\frac{\partial f_i(\Delta \hat{c}^i)}{\partial \Delta \hat{c}^i} \ge 0$, 
to ensure that higher attribute difference never decreases the overall face distance.
We implement GAM as neural layers using Node-GAM~\cite{chang2021node} (Fig.~\ref{fig:architecture}.4).

\subsection{AlignFace2AFC Perception Metric Learning}
We define \rev{AlignFace2AFC}, a multi-task triplet metric for similarity perception of a 
% We formulate the task as a multi-task triplet metric learning 
face triplet $\bm{x} = \langle x_A, x_\text{Ref}, x_B \rangle$ (Fig.~\ref{fig:architecture_triplet}).
For each triplet,
we apply AlignFace twice to compare face pairs $\langle x_A, x_\text{Ref} \rangle$ and $\langle x_B, x_\text{Ref} \rangle$ separately to predict their pairwise distances and calculate the \textbf{Relative Distance Difference} for the overall face ($\Delta \hat{d} = \hat{d}_{BR} - \hat{d}_{AR}$) and attributes ($\Delta\Delta \hat{c} = \Delta \hat{c}_{BR}$ - $\Delta \hat{c}_{AR}$).

We collected human ground-truth labels of similar face selection for overall face $y_d \in \{+1, -1\}$ and per-attribute $y_c^i \in \{+1, -1\}$.
Each $y$ is +1 if face B $x_B$ is selected as more similar to the reference face $x_\text{Ref}$, and -1 if face A $x_A$ is selected instead.
Note that the similarity label has the opposite sense of the distance that AlignFace predicts, so we need to flip it to \textbf{Dissimilarity Label} $-y_d$ and $-\bm{y}_c$.
AlignFace is trained by minimizing the squared hinge loss~\cite{fu2023dreamsim} between the dissimilarity labels and relative distance differences, i.e.,
\begin{equation}
\mathcal{L}^i_c(\Delta \Delta \hat{c}_i, y_c^i)
= \left(\max\big(0,\, m - \Delta \Delta \hat{c}_i \cdot (-y_c^i) \big)\right)^2,
\label{eq:attr}
\end{equation}
\begin{equation}
\mathcal{L}_d(\Delta \hat{d}, y_d)
= \left(\max\big(0,\, m - \Delta \hat{d} \cdot (-y_d) \big)\right)^2,
\label{eq:face}
\end{equation}
where $m$ is a margin (set to $m = 0.05$), to enforce relative ranking via attract-and-repelling. 

To manage training instability, we first train the attribute-based difference prediction with per-attribute loss $\mathcal{L}^i_c$, freeze it, and then train the Neural GAM with overall distance loss $\mathcal{L}_d$.

\section{Experiments}
We evaluated AlignFace 
% through modeling experiments to demonstrate how well it aligns with human judgments and reasoning processes.
to investigate its alignment and attribution faithfulness to human perception, compared to baselines.

\subsection{Experimental Settings}
\subsubsection{Implementation Training Details}
% \outline{
% - Implement AlignFace in PyTorch.
% - Rational: the large number of parameters of models, vs. relative small among of perceptual data.
% - Frozen textual encoder.
% - (Low-Rank Adaptation) LoRA to finetune the pretrained visual encoder.
% -- Report the hyperparameters ($r=?$, dropout $p=?$, $a=?$)...
% -- Which optimizer?
% % -- 1st stage vs. 2nd stage
% - Train CBM from scratch.
% - Computational resources: a server equipped with 8 NVIDIA RTX 3090 GPUs
% }

We implemented AlignFace in PyTorch, freezing the text encoder and adapting the visual encoder with LoRA~\cite{hu2022lora} ($r=4$, $\alpha=8$, dropout $=0.5$). The fusion module comprises cross-attention followed by a three-layer MLP (dropout $=0.2$), while NodeGAM~\cite{chang2021node} uses 50 knots. We trained both attribute- and face-level predictions with squared hinge loss (margin $=0.05$) using Adam (learning rate $=10^{-3}$, weight decay $=10^{-4}$, batch size $=16$) on NVIDIA RTX 3090 GPUs. 
% ---------- Runtime/cost analysis (AC) exact wording ----------
% r5Tm's comments: "Additionally, including runtime analysis and computational cost would be beneficial, particularly given the multiple architectural components integrated into the proposed framework.”
% We didn't respond to this.
% \rebuttalpoint{Runtime/cost analysis (AC).}
\rev{AlignFace has 29.42M trainable parameters and required 157.16ms per face pair on a single RTX 3090 GPU.}

\subsubsection{Evaluation Criteria}
% To assess the alignment of various perceptual similarity metrics with human judgment, we evaluate face triplets by emulating the 2AFC task on our held-out test set.
% For each metric, we derive a binary preference based on relative similarities: $y=1$ if $d_{AR} < d_{BR}$, and $y=0$ otherwise.
% by comparing the relative similarity between pairs $y=1 \Rightarrow s(A, R) \succ s(B, R)$.
To assess alignment with human 2AFC judgments, we evaluate each metric on held-out triplets by comparing its predicted preference with human annotations.
Given the relative distance difference $\Delta \hat{d}$ and human labels $y_d \in \{+1,-1\}$ defined earlier, we compute agreement as:
\[
a = \mathbf{1}[\Delta \hat{d} \cdot (-y_d) > 0].
\]
We evaluate alignment from two perspectives:
\begin{enumerate}[label=\arabic*), leftmargin=*, itemsep=0em, topsep=0.2em, parsep=0pt, partopsep=0pt]
    \item \textbf{Behavioral Agreement}: We calculate the agreement score between averaged human ratings and binarized model preferences.
    \item \textbf{Reasoning Analysis}: To gain deeper insights into the model behavior, we visualize and analyze partial dependence plots (PDPs) to compare the relationship between individual attribute similarities and overall face perception. 
\end{enumerate}
\begin{figure}[!t]
    \centering
    \vspace{-0.2cm}
    \includegraphics[width=1\linewidth]{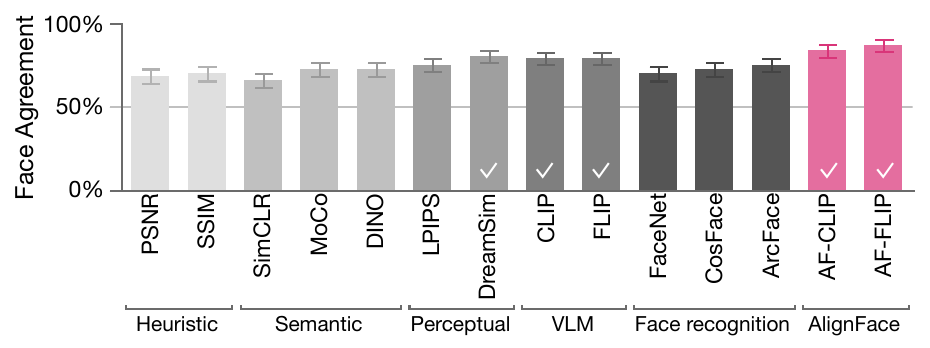}
    \vspace{-0.2cm}
    \caption{
    Model-human agreement on overall face perception. 
    The grey line represents random guessing. 
    Error bars show 95\% confidence intervals. 
    % ---------- Baseline comparison (r5Tm) exact wording ----------
    % Reply: "We compared against FaceNet, CosFace, and ArcFace baselines, finding that AlignFace is significantly better (see Fig.~R.\ref{fig:face_alignment_fr}). This could be due to their weaker training on smaller datasets without VLM pretraining."
    % \rebuttalpoint{Experimental comparison against face-specific baselines (r5Tm).}
    \rev{Baselines without $\checkmark$ are significantly lower than AF-FLIP (Dunnett’s test, $\alpha=0.05$).}
    }
    \vspace{-0.4cm}
    \label{fig:results_alignment_face}
\end{figure}
\begin{figure*}[!t]
    \centering
    \vspace{-0.2cm}
    \includegraphics[width=0.95\linewidth]{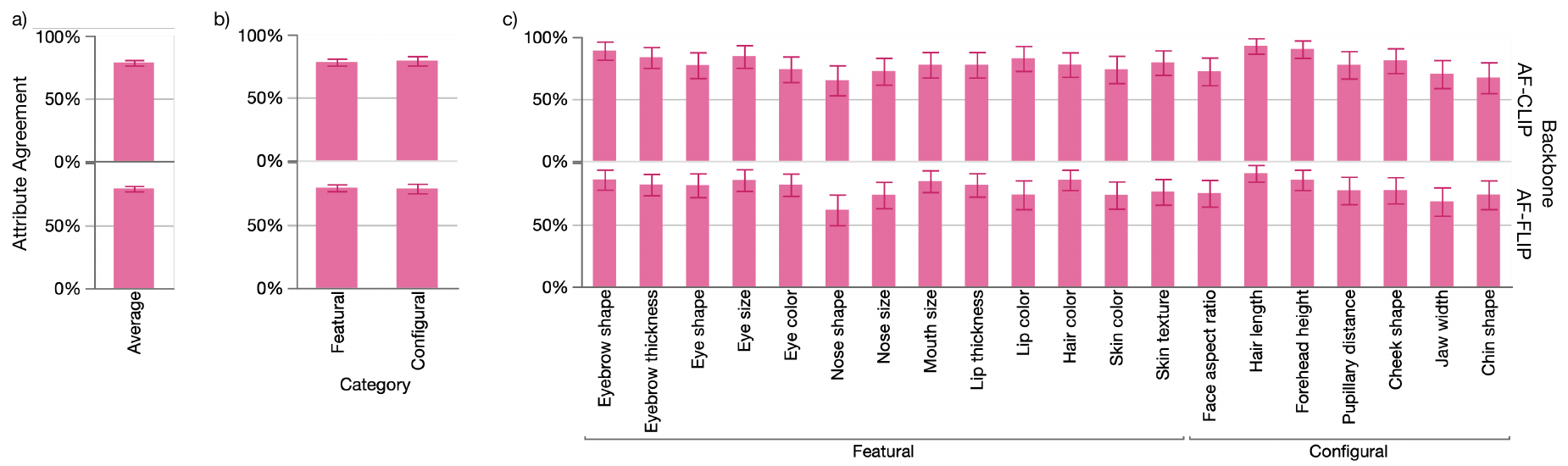}
    \vspace{-0.2cm}
    \caption{
        % \todo{make figure 6b here, get rid of grey bars.}
        Attribute perception agreement of a) averaged across all attributes, b) grouped into featural and configural categories, and c) at individual-level attributes.
        % Pretrained CLIP or FLIP with a randomly weighted fusion module as baselines. 
    } 
    \vspace{-0.2cm}
    \label{fig:results_alignment_attr}
\end{figure*}
\subsubsection{Baseline Comparators}
\label{sec:baselines}
For overall face similarity, we compare AlignFace against widely used metrics for assessing generative content quality. These include:
\begin{enumerate}[label=\arabic*), leftmargin=*, itemsep=0em, topsep=0.2em, parsep=0pt, partopsep=0pt]
    \item Heuristic metrics: PSNR~\cite{hore2010image} and SSIM~\cite{wang2004image}.
    \item Self-supervised semantic visual representation models: SimCLR~\cite{chen2020simple}, MoCo~\cite{he2020momentum} and DINOv2~\cite{oquab2024dinov2}.
    \item Visual-language models: CLIP~\cite{radford2021learning}, FLIP~\cite{li2024flip}.
    \item Perceptual similarity models: LPIPS~\cite{zhang2018unreasonable} and DreamSim~\cite{fu2023dreamsim}.
    % ---------- Rebuttal (r5Tm), exact wording ----------:
    % Experimental comparison against face-specific baselines (r5Tm). 
    % Reply: "We compared against FaceNet, CosFace, and ArcFace baselines, finding that AlignFace is significantly better (see Fig. R.1)."
    \item 
    % \rebuttalpoint{Experimental comparison against face-specific baselines (r5Tm).} 
    \rev{Face-specific models: FaceNet~\cite{schroff2015facenet}, CosFace~\cite{wang2018cosface} and ArcFace~\cite{deng2019arcface}.}
\end{enumerate}

% Since these standard metrics are generally unable to capture the granular facial semantics required for attribute-level comparison, we establish an attribute-level baseline using CLIP and FLIP~\footnote{Attribute-level evaluation using CLIP and FLIP with pretrained weights.}. 
We also performed ablation studies to examine the benefits to overall and per-attribute alignment of the modules of AlignFace.
For these comparisons, we initialize our fusion module with randomized parameters to provide a rigorous point of reference.

\subsection{Alignment Evaluation}
\label{sec:main_evaluation}
% \outline{
% - 1. Our AlignFace is better in both overall face perception alignment. (response) \\
% - 2. We also provide alignment capability for facial attribute perception.
% - 3. The GAM PDP plots show alignface not only more interpretable, but also align with cognitive process (reasoning). \\
% - 4. Show case how others make inconsistent rating with humans, while we are aligned?
% }

% Fig.~\ref{fig:results_alignment_face} compares similarity evaluation under 2AFC tasks for both overall face perception and attribute-level perception.
\begin{figure}[!t]
    \centering
    \vspace{-0.0cm}
    \includegraphics[width=1.0\linewidth, trim=0.05cm 0cm 0cm 0cm, clip=true]{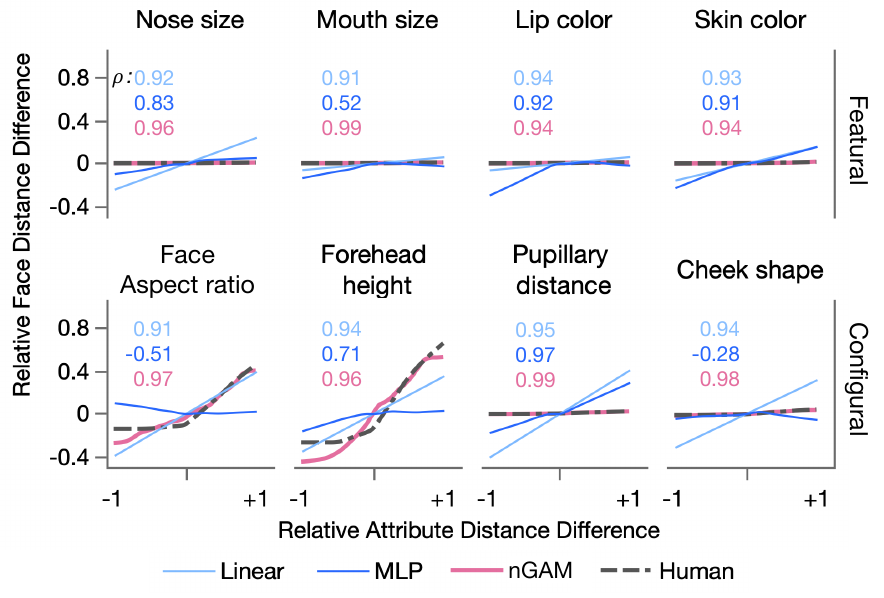}
    \vspace{-0.5cm}

    % ---------- Rebuttal (jvsS), exact wording ----------:
    % Verify reasoning about relevant attributes (jvsS). 
    % Reply: "We clarify the reported high agreements (Mean \rho = 0.98) between AlignFace's Neural GAM and human PDP, indicating faithful reasoning."
    % \rebuttalpoint{Verify reasoning about relevant attributes (jvsS).} 
    \caption{Partial dependence plots (PDPs) of selected attributes, comparing human judgments with Linear, MLP and nGAM. 
    The numbers in graphs are Pearson correlations $\rho$ between each model and humans.
    \rev{High correlations for nGAM (0.94--0.98) indicate strong human alignment of AlignFace.}
    }
    \vspace{-0.2cm}
    \label{fig:results_alignment_GAM}
\end{figure}
\begin{figure}[!t]
    \centering
    % \vspace{-0.2cm}
    \includegraphics[width=0.95\linewidth,
    trim={0 0.1cm 0 0},
    clip]{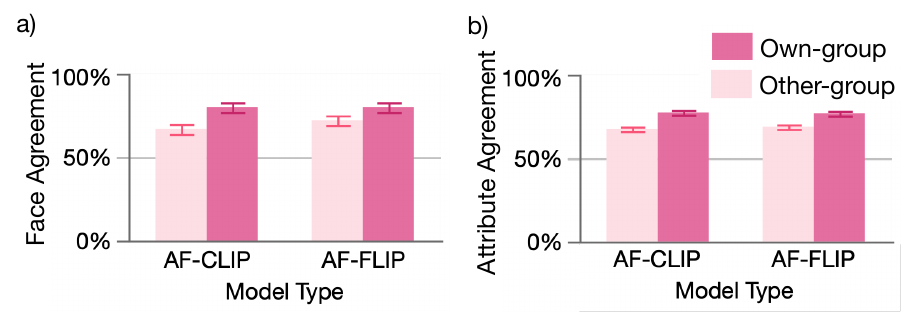}
    \vspace{-0.2cm}
    \caption{Cross-group perception agreement evaluations on
    a) face and b) averaged attributes.
    }
    \vspace{-0.2cm}
    \label{fig:cross_group}
\end{figure}
\textbf{Overall Face Perception Alignment.}
Heuristic approaches (PSNR, SSIM) and self-supervised representation models (SimCLR, MoCo, DINO) exhibit the lowest consistency with human ratings (Fig.~\ref{fig:results_alignment_face}a).
In contrast, perceptual methods outperform both groups, with DreamSim achieving slightly higher agreement than LPIPS.
Surprisingly, approaches based on pretrained contrastive visual-language models (VLMs) achieve relatively strong performance among the baselines.
% ---------- Baseline comparison (r5Tm) exact wording ----------
% Reply: "We compared against FaceNet, CosFace, and ArcFace baselines, finding that AlignFace is significantly better (see Fig.~R.\ref{fig:face_alignment_fr}). This could be due to their weaker training on smaller datasets without VLM pretraining."
% \rebuttalpoint{Experimental comparison against face-specific baselines (r5Tm).}
\rev{FaceNet, CosFace, and ArcFace baselines performed significantly worse, likely due to their smaller training datasets and lack of VLM pretraining.}
Overall, AlignFace outperforms all competing methods.
We also observed additional gains when using domain-specific encoders (i.e., AlignFace (FLIP) vs.\ AlignFace (CLIP)), suggesting that FLIP better captures facial semantics.

\textbf{Face Attributes Alignment.}
% Since baseline metrics do not support explicit attribute-level comparisons, we construct two baseline conditions using CLIP and FLIP backbones, each augmented with a multi-modal fusion module $F_z$ initialized with random weights. 
% As shown in Fig.~\ref{fig:results_alignment_face}b, \todo{these baselines perform no better than random guessing (i.e., $50\%$ agreement), confirming their inability to capture attribute-level perceptual similarity.}
% In contrast, AlignFace substantially improves alignment, with consistent gains across backbones, indicating robustness to the choice of visual encoder.
% Beyond the overall improvement, consistent gains are observed across all face attributes (Fig.~\ref{fig:results_alignment_face}b).
% However, the magnitude of improvement varies by attribute. Attributes associated with larger or more visually salient regions (e.g., forehead height, hair color) show greater gains than attributes defined by finer details (e.g., nose shape, lip thickness). 
% Finally, we did not observe a significant difference between featural and configural attributes, suggesting that AlignFace benefits both types of representations similarly.
% ---------- Updated results after new participants ----------:
We investigate how well AlignFace aligns with human perception across face attributes\footnote{\rev{As a mediation check, 
we also validated that general-purpose VLMs could accurately encode objectively-verifiable, facial attributes by comparing AlignFace's predicted attribute sizes and distances against ground-truth heuristic inter-landmark measures~\cite{lugaresi2019mediapipe}. 
Appendix Table~\ref{tab:manipulation_check} shows high correlations (Median $\rho = 0.947$).}}.
Fig.~\ref{fig:results_alignment_attr}a shows that AlignFace had reasonably high attribute agreement (\rev{$M = 0.79$} for FLIP backbone). This was lower than the overall face agreement, perhaps due to attribute diversity and data sparsity.
% \toreview{AlignFace could better align for configural attributes than featural ones (Fig.~\ref{fig:results_alignment_attr}b).}
Specifically, it had the highest agreement for Hair length, \rev{Eyebrow shape and Forehead height,} and the lowest agreement for \rev{Nose shape, Jaw width,} and Chin shape (Fig.~\ref{fig:results_alignment_attr}c).

% \textbf{Cognitive Reasoning Alignment.} 
\textbf{Attribute Influence Alignment.} 
Next, we examined whether AlignFace was ``right for the right reason''~\cite{ross2017right} by comparing its attribute influence on overall perception against that of human perception.
Specifically, we compared the individual GAM function shapes of AlignFace due to its Attribute-Influence Neural GAM module (Section~\ref{sec:GAM_module}) represented in the triple-based dual-pair comparison (Fig.~\ref{fig:architecture_triplet}) against the human GAM (Section~\ref{sec:user-study}, Fig.~\ref{s_fig:gam_alignment_full}).
Fig.~\ref{fig:results_alignment_GAM} shows how well AlignFace's partial dependence plots align with human perception relations, achieving high Pearson correlations $\rho$.
See Appendix Fig.~\ref{s_fig:gam_alignment} for all attributes.
% Specifically, we fit a GAM with the same architecture as AlignFace using human-provided attribute and face annotations collected in Sec.~\ref{sec:user-study}, thereby approximating the aggregation process underlying human reasoning.
% We then compared the resulting partial dependence plots (PDPs) across humans, AlignFace, Linear, and MLP, as shown in Fig.~\ref{fig:results_alignment_GAM} (see Appendix Fig.~\ref{s_fig:gam_alignment} for all attributes).
% AlignFace shows strong consistency with human trends and achieves the highest Pearson correlation ($\rho$).
% \todo{In contrast, the MLP exhibits inconsistent and sometimes reversed trends relative to human perception, leading to substantially lower correlation.
% The linear model maintains a reasonable correlation due to its monotonic form, but fails to capture the nonlinear patterns observed in human judgments.
% These results indicate that AlignFace not only matches human decisions, but also more faithfully captures the underlying reasoning process.}

% \subsection{Improved alignment with subpopulation tuning}
\textbf{Subpopulation Alignment.}
% \outline{
% - To support face perception metric aligned with subpopulation perceptions, we tune AlignFace and show benefits under own group. \\
% - Improved alignment for within group }
% \outline{
% - AlignFace supports cross dataset evaluation: train on one dataset and evaluation on the other ... \\
% - no significant different when evlauating cross-domain faces.
% }

% \textbf{Improved alignment with subpopulation tuning.}
Given the own-group bias observed 
% \rebuttalpoint{Scoping own-group claims to White/Asian comparison (AC).}
\rev{for White and Asian participants} in Section~\ref{sec:user-study}, we next investigated its impact on AlignFace.
Fig.~\ref{fig:cross_group} shows that AlignFace is in good agreement with human overall perception across the faces of the Own and Other-groups.
Notably, agreement was stronger for Own-group perception than Other-group, perhaps because the model was trained on fewer Other-group faces due to imbalanced ethnicities in the face datasets.
% To this end, we partitioned the dataset into own-group and other-group subsets, retrained two AlignFace models on each subset, and evaluated them on both in-group and cross-group held-out test sets.
% The results in Fig.~\ref{fig:cross_group} show that AlignFace achieves a superior agreement with human ratings when evaluated using an own-group model compared with an other-group model.
% This pattern holds for both face-level (Fig.~\ref{fig:cross_group}a) and attribute-level (Fig.~\ref{fig:cross_group}b) evaluations. 
See Appendix Fig.~\ref{s_fig:cross_group_attr} for results across all attributes.
This finding suggests that the proposed approach benefits from tailoring the model to specific subpopulations.

% \subsection{Generaliability across face datasets.}
\subsection{Cross-Dataset Generalizability}
% \textbf{}
\begin{table}[!t]
\centering
\vspace{-0.2cm}
\caption{
    Agreement under within- and cross-dataset evaluation for AF-CLIP and AF-FLIP.
    Consistent performance indicates learned generalized perception across datasets. 
}
\vspace{-0.2cm}
\small
\setlength{\tabcolsep}{6pt} 
\renewcommand{\arraystretch}{0.9}   
\begin{tabular}{r cc cc}
\toprule
\multirow{2}{*}{\textbf{Setting}}
& \multicolumn{2}{c}{\textbf{AF-CLIP}} 
& \multicolumn{2}{c}{\textbf{AF-FLIP}} \\
\cmidrule(lr){2-3} \cmidrule(lr){4-5}
& Face & Attr & Face & Attr \\
\midrule
% Within-dataset & \textbf{78.4}\% & \textbf{78.4}\% & 73.0\% & \textbf{78.7}\% \\
Within-dataset & 79.9\% & 77.9\% & 81.3\% & 78.3\% \\
Cross-dataset  & 79.4\% & 72.9\% & 81.3\% & 73.2\% \\
\bottomrule
\end{tabular}
\label{tab:cross_dataset}
\vspace{-0.4cm}
\end{table}
% Following a similar procedure, w
We investigated the impact of dataset domain shifts in face images on AlignFace's performance.
To this end, we retrained two AlignFace models, with CLIP and FLIP backbones, using targets from the CMU Multi-PIE~\cite{gross2010multi} and CelebA~\cite{liu2015faceattributes} datasets separately, and evaluated them under both within-dataset and cross-dataset settings.
Table~\ref{tab:cross_dataset} shows that AlignFace achieves consistent agreement scores across both settings.
%, with no significant differences observed (
See Appendix Fig.~\ref{s_fig:cross_dataset_attr} for per-attribute results.
This result 
% suggests that the influence of out-of-domain data on the proposed metric is limited, indicating 
indicates strong robustness and generalizability across datasets, 
% perhaps due to the open-domain world knowledge of the VLM, grounding of using cognitively-relevant attributes, and alignment of relational trends between attributes and overall perception.
perhaps due to the VLM's open-domain knowledge, the use of cognitively-grounded attributes, and the alignment of relationships between attributes and overall perception.
% A likely explanation is that, during face triplet learning, we adopt a comparative metric learning framework that models relative ranks between pairs rather than relying on absolute predictions, as in conventional classification or regression approaches. 
% This property allows our metric to be reliably applied to face data under unknown domain shifts.

\subsection{Ablation Studies}
\begin{figure}[!t]
    \centering
    \vspace{-0.2cm}
    \includegraphics[width=1\linewidth,
    trim={0 0.1cm 0 0},
    clip]{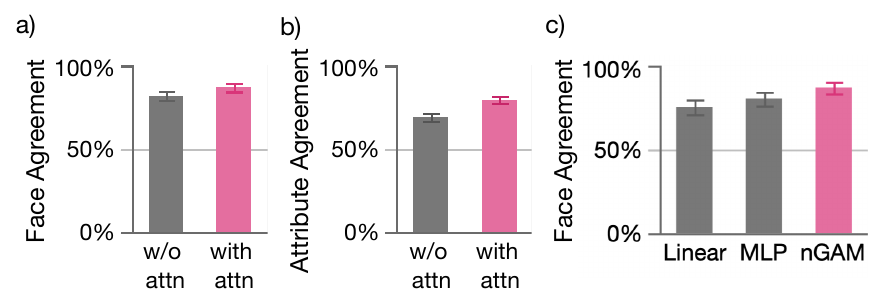}
    \vspace{-0.5cm}
    \caption{
    Results of ablation studies on Attribute-Gated Cross-Attention: a) overall face model-human agreement, b) average attribute agreement; 
    and on Attribute-Influence Neural GAM: c) overall face agreement.
    }
    \vspace{-0.4cm}
    \label{fig:results_ablation}
\end{figure}

% Prior research has raised concerns that providing interpretability may degrade a model's performance~\cite{rudin2019stop}.
We conducted ablation studies to evaluate the contribution of two key components in AlignFace: 
1) Attribute-Gated Cross-Attention for attribute-conditioned representation learning, and 
2) Attribute-Influence Neural GAM (nGAM) for aggregating attribute differences into the overall face distance perception.
Results are in Fig.~\ref{fig:results_ablation}.

\textbf{Attribute-Gated Cross-Attention.}
We replaced the Attribute-Gated Cross-Attention module with a direct fusion variant, where the image-pair and attribute embeddings are concatenated and fed into an MLP.
% , removing attribute-conditioned attention.
Fig.~\ref{fig:results_ablation}a,b show a clear drop in agreement with human perception at both the overall and attribute levels, with a greater degradation for attribute-level agreement.
Even with supervision from human labels, the direct fusion variant fails to capture attribute-specific perceptual cues as effectively.
These results highlight the importance of attribute-conditioned visual-semantic interaction for modeling human-aligned similarity.

\textbf{Attribute-Influence Neural GAM.}
We further replaced the Neural GAM module with two alternative aggregation heads: a Linear model and a Multi-Layer Perceptron (MLP).
The Linear model assumes a simple additive weighting across attributes, 
while the MLP captures nonlinear interactions in a black-box manner.
Fig.~\ref{fig:results_ablation} shows that the Linear model had the lowest agreement, indicating that simple additive weighting is insufficient to model perceptual similarity.
The MLP achieved performance comparable to Neural GAM, but lacked interpretability.
% In contrast, Neural GAM maintains strong agreement with human judgments while providing intrinsic interpretability, demonstrating that structured, interpretable aggregation does not compromise performance.
%
We also examined the attribute-influence trends using partial dependence plots for the ablated models.
Fig.~\ref{fig:results_alignment_GAM} shows the poor fit of the Linear model and MLP.
Notably, the complexity of the MLP did not help its agreement compared to Linear. 
For example, MLP learned spurious trends of decreasing overall perceived difference with increasing relative attribute distance difference for Cheek shape.
Hence, Neural GAM had the highest human alignment while being interpretable.
% \textbf{Neural GAM.}
% In particular, we replaced the Neural GAM module in AlignFace (Sec.~\ref{sec:GAM_module}) with two alternatives (Linear \& MLP).
% \begin{enumerate}[label=\arabic*), leftmargin=*, itemsep=0em, topsep=0.2em, parsep=0pt, partopsep=0pt]
%     \item \textbf{Linear Model:} $F(\Delta \hat{\bm{c}}) = \sum_{i=1}^{N} w_i \Delta \hat{c}^i + \beta$. This serves as a naive baseline, assuming a linear relationship where each attribute contributes proportionally to the overall perception.    
%     \item \textbf{Multi-Layer Perceptron (MLP):} A nonlinear model $F(\Delta \hat{\bm{c}})$ that captures complex, high-order interactions among attributes. While highly expressive, MLPs are prone to modeling spurious correlations and lack interpretability.
% \end{enumerate}
% \todo{Results in Fig.~\ref{fig:results_ablation} demonstrate that ...
% It reveals that with ante-hoc interpretability, AlignFace achieves comparable agreement with human perception.}
% These findings support the design of AI models with intrinsically interpretable mechanisms, where insights derived from human cognition potentially guide and augment model behavior and bring extra benefits.

% \outline{
% - ablate on Attribute-Gated cross attention:
% -- 1. Full, with-sinp 
% - We ablate on Neural GAM: \\
% -- 1. Linear, 2. MLP, 3. GAM  \\
% -- Show similar performance but GAM with the best interpretability.}

% =========================
\section{Discussion}
% \note{
% - Limitations \\
% - 1: limited size of dataset. \\
% - 2: human annotators highly bias towards Caucasian. \\
% - 3: as a preliminary exploration on human-aligned face similarity metric, not validation on more GAN synthesis scenarios: privacy, retouching, makeup-transfer, restoration, enhancement... \\

% - Discussion points \\
% - 1: our approach of CBM + GAM to support semantic-aware visual similarity can generalize to other visual/audio tasks. \\
% - 1.1: for general image task, semantic-aware visual similarity can facilitate image retrieval based on visual patterns rather than text queries. \\
% - 1.2: for audio task, could support music retrieval based on humming. \\
% - 2: our philosophy of achieving human-aligned modeling at both behavioral responses and cognitive reasoning. \\
% }
% \todo{emphasizing: attribute relevance, nonlinear perceptual scaling, and demographic variance}
% ---------- Rebuttal (r5Tm), exact wording ----------:
% Reply: "Thank you, we have revised accordingly."
% \rebuttalpoint{Emphasize cognitive characteristics more clearly in contributions and discussion (r5Tm).}
% \toreview{A key contribution of AlignFace is its cognitively grounded modeling of face similarity. 
% Beyond predicting human judgments, it captures the relevance of various facial attributes, nonlinear effects on perceived similarity, and demographic variation. 
% By integrating attribute concepts with a neural GAM, AlignFace interprets both human responses and their underlying reasoning.}
\rev{We have shown the benefits of grounding face similarity modeling with the cognitive principles of featural and configural attributes, nonlinear scaling, and own-group bias.}
While this work establishes a foundation for human-aligned face similarity, it focuses on controlled perceptual settings with a relatively compact dataset and a specific participant demographic,
\rebuttalpoint{Over-filtration of data (JeJP).}
\rev{and excludes ambiguous trials with 30--70\% rater agreement\footnote{\rev{Although common in crowdsourced model training~\cite{zhang2018unreasonable, fu2023dreamsim}, this may omit Point of Subjective Equality (PSE) ``hard examples'' for fine-grained perceptual differences.}}.}
% Extending the framework 
Future work could extend it 
to broader populations and more diverse generative scenarios, such as privacy-preserving obfuscation, makeup transfer, and restoration.
% , is a natural next step.

% ---------- Rebuttal (JeJP), exact wording ----------:
% Reply: "We clarify that we followed common practice in model training from crowd-labeling~\cite{zhang2018unreasonable, fu2023dreamsim} to exclude ambiguous trials (30\%--70\% agreement) to ensure high-quality training labels. We have acknowledged this limitation in the Discussion section that this could omit PSE ``hard examples'' for fine-grained perception."
% \rebuttalpoint{Over-filtration of data (JeJP).}
% \toreview{In addition, our findings are based on the high-quality instances, excluding ambiguous trials with 30\%-70\% rater agreement. 
% Although common in crowdsourced model training~\cite{zhang2018unreasonable, fu2023dreamsim}, this may omit challenging examples involving fine-grained perceptual differences.}

Beyond faces, AlignFace suggests a general approach for semantic-aware visual similarity.
% that can be extended to other domains.
Although we focus on attributes for face perception, the prompt-based VLM can model open-ended domain concepts, such as ABCD criteria for assessing skin-lesion disease progression~\cite{abbasi2004early}.
% Although we had identified grounded attributes for face perception, our approach models concepts via prompt-based VLM, so other domains can be modeled using open-ended textual descriptions (e.g., comparing skin lesion disease progression due to ABCD criteria~\cite{abbasi2004early}).
% For instance, it may facilitate image retrieval based on complex visual content rather than text queries~\cite{jain1996image, wan2014deep} and support audio tasks like humming-based music retrieval~\cite{ghias1995query, murthy2018content}. 
Ultimately, our method advances toward human-aligned modeling that mirrors both behavioral responses and underlying cognitive reasoning, which has the potential to inspire other work on human-AI alignment across diverse perceptual tasks.

% =========================
\section{Conclusion}
% In this paper, w
We have presented AlignFace, a human-aligned face similarity metric to model both face and fine-grained attribute perception.
Grounded in scientific cognitive principles, it achieves alignment across both predictive behavioral responses and underlying model reasoning.
Experimental evaluations demonstrate that AlignFace significantly outperforms existing domain-free metrics in mirroring human judgment.
We further provide an in-depth analysis of subpopulation influences, dataset-driven domain shifts, and the architectural impact.  
Ultimately, this work establishes a more faithful proxy for human face perception, providing a reliable foundation for evaluating generative facial content.

\newpage
% =========================
% Acknowledgements (camera-ready only)
% =========================
\begin{acks}
\rev{This research is supported by 
the Ministry of Education, Singapore (Award No: T2EP20121-0040),
the National Research Foundation, Singapore and Infocomm Media Development Authority under its Trust Tech Funding Initiative (Award No: DTC-RGC-09), 
a Google Research Scholar Award, and 
the NUS Institute for Health Innovation and Technology (iHealthtech).
% Any opinions, findings and conclusions or recommendations expressed in this material are those of the author(s) and do not reflect the views of the Ministry of Education, Singapore, National Research Foundation, Singapore, Infocomm Media Development Authority, Google, and iHealthtech.
The views expressed are those of the authors and do not necessarily reflect those of the funders.}
\end{acks}

\bibliographystyle{ACM-Reference-Format}
\balance
\bibliography{references}

\appendix
\captionsetup[figure]{labelfont={bf},font={small},name={Appendix Fig.},labelsep=period}
\captionsetup[table]{labelfont={bf},font={small},name={Appendix Table},labelsep=period}

\clearpage
\onecolumn
\section{Appendix}
\subsection{Dataset and Participant Details}
\label{sec:user_study_details}
\begin{table}[!ht]
\centering
\small
\caption{Demographic distribution of face targets selected during dataset curation.}
\renewcommand{\arraystretch}{0.8}
\vspace{-0.3cm}
\begin{tabular}{lcccc}
\toprule
\textbf{Ethnicity} & \multicolumn{2}{c}{White} & \multicolumn{2}{c}{Asian}   \\
\midrule
\textbf{Gender} & Female & Male & Female & Male \\
\midrule
\textbf{Multi-PIE} & 20 & 20 & 10 & 10 \\
\textbf{CelebA} & 20 & 20 & 10 & 10 \\
\bottomrule
\end{tabular}
% \vspace{-0.3cm}
\label{s_table:dataset_distro}
\end{table}

\begin{figure}[!ht]
    \centering
    \includegraphics[width=0.8\linewidth]{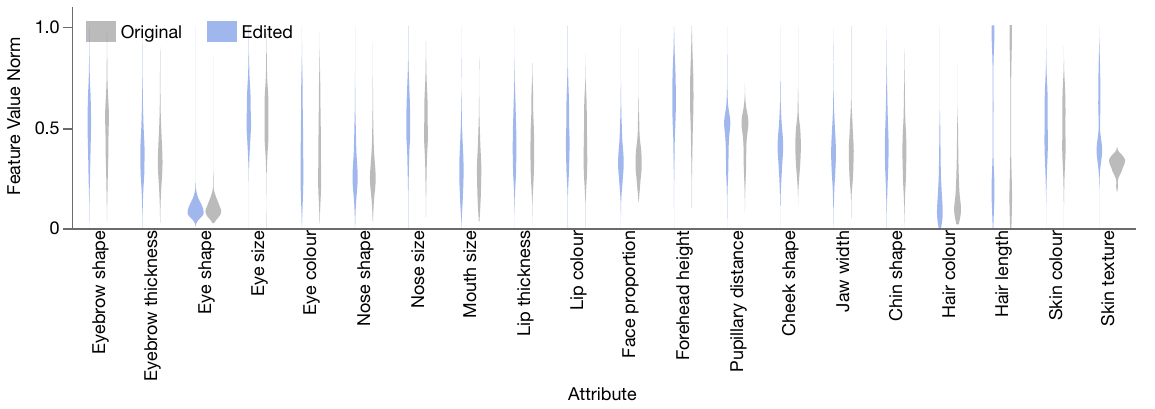}
    \vspace{-0.3cm}
    \caption{Statistical distribution of feature values for source and edited faces across 20 face attributes. These feature values are computed using geometry-based heuristic methods and serve as proxy evaluation metrics.}
    \vspace{-0.3cm}
    \label{s_fig:stats_edit_face}
\end{figure}

% ---------- Updated results after new participants ----------:
\begin{table}[!ht]
\small
\caption{Demographic statistics of participants in our psychophysical similarity measurement studies.}
\begin{tabular}{rcccccccccc}
\hline
& \multicolumn{5}{c}{\textbf{Ethnicity}}   & \multicolumn{3}{c}{\textbf{Gender}} & \multicolumn{2}{c}{\textbf{Age}} \\ \hline
& White & Asian  & Black& Hispanic & Other & Female      & Male      & Other     & Range          & Median          \\ \hline
Face perception      & 35    & \rev{35}     & 2    & 2        & \rev{4}     & \rev{38}          & \rev{38}        & \rev{2}         & \rev{21-78}          & \rev{44.0}            \\ \hline
Attribute perception & \rev{342}   & \rev{187}      & \rev{53}  & 9        & 20    & \rev{324}         & \rev{286}       & 1         & \rev{21-84}          & \rev{44.0}            \\ \hline
\end{tabular}
% \vspace{-0.3cm}
\label{s_table:participants} 
\end{table}
\newpage

\subsection{Statistical Analysis}
\begin{table}[!h]
\centering
\small
% ---------- Rebuttal (r5Tm), exact wording ----------:
% Inter-rater agreement and significance statistics (r5Tm). 
% Reply: "We have added Fleiss' kappa statistics to quantify inter-rater agreement. The overall Fleiss' kappa is 0.604, indicating substantial agreement among annotators. We also include significance analysis using Dunnett's test."
\caption{
% \rebuttalpoint{Inter-rater agreement (r5Tm).} 
\rev{Inter-rater agreement (Fleiss' $\kappa$)~\cite{landis1977measurement} for overall face and 20 facial attributes.}
% Items are (face triplet, attribute) pairs rated via binary 2AFC.
}
\setlength{\tabcolsep}{6pt} 
\renewcommand{\arraystretch}{0.9}
\begin{tabular}{rcc}
\toprule
\textbf{Attribute} & \textbf{Fleiss' $\kappa$} & \textbf{Agreement} \\
\midrule
Overall & 0.604 & Moderate \\
\midrule
% Hair length         & 0.736 & Substantial \\
% Eyebrow shape       & 0.718 & Substantial \\
% Eye color           & 0.705 & Substantial \\
% Eyebrow thickness   & 0.703 & Substantial \\
% \midrule
% Forehead height     & 0.613 & Moderate \\
% Eye size            & 0.605 & Moderate \\
% Nose size           & 0.604 & Moderate \\
% Nose shape          & 0.598 & Moderate \\
% Face aspect ratio   & 0.598 & Moderate \\
% Lip thickness       & 0.595 & Moderate \\
% Jaw width           & 0.594 & Moderate \\
% Skin texture        & 0.578 & Moderate \\
% Mouth size          & 0.577 & Moderate \\
% Cheek shape         & 0.576 & Moderate \\
% Eye shape           & 0.560 & Moderate \\
% Chin shape          & 0.546 & Moderate \\
% Hair color          & 0.544 & Moderate \\
% Lip color           & 0.538 & Moderate \\
% Skin color          & 0.537 & Moderate \\
% Pupillary distance  & 0.457 & Moderate \\
Eyebrow shape & 0.718 & Substantial \\ Eyebrow thickness & 0.703 & Substantial \\ Eye shape & 0.560 & Moderate \\ Eye size & 0.605 & Moderate \\ Eye color & 0.705 & Substantial \\ Nose shape & 0.598 & Moderate \\ Nose size & 0.604 & Moderate \\ Mouth size & 0.577 & Moderate \\ Lip thickness & 0.595 & Moderate \\ Lip color & 0.538 & Moderate \\ Hair color & 0.544 & Moderate \\ Skin color & 0.537 & Moderate \\ Skin texture & 0.578 & Moderate \\ \midrule Face aspect ratio & 0.598 & Moderate \\ Hair length & 0.736 & Substantial \\ Forehead height & 0.613 & Moderate \\ Pupillary distance & 0.457 & Moderate \\ Cheek shape & 0.576 & Moderate \\ Jaw width & 0.594 & Moderate \\ Chin shape & 0.546 & Moderate \\ 
\bottomrule
\end{tabular}
\label{tab:fleiss_kappa}
\end{table}

\begin{table}[!h]
\centering
\small
% ---------- Rebuttal (r5Tm), exact wording ----------:
% Significance analysis (r5Tm). 
% Reply: Dunnett's test (\alpha = 0.05) shows AF-FLIP has the highest mean performance, which is significantly higher than most baselines (Fig. R.1, Appendix Table A.6)."
\caption{
% \rebuttalpoint{Significance analysis (r5Tm).} 
\rev{Dunnett's test comparing AF-FLIP against baselines ($\alpha = 0.05$).}}
\setlength{\tabcolsep}{6pt} 
\renewcommand{\arraystretch}{0.9} 
\label{tab:dunnett}
\begin{tabular}{r c c c}
\hline
Method & Comparison & Diff. & $p$-value \\
\hline
PSNR     & AF-FLIP & -0.184642 & $<.0001$ \\
SSIM     & AF-FLIP & -0.171484 & $<.0001$ \\
SimCLR   & AF-FLIP & -0.210958 & $<.0001$ \\
MoCo     & AF-FLIP & -0.145168 & $<.0001$ \\
DINO     & AF-FLIP & -0.145168 & $<.0001$ \\
LPIPS    & AF-FLIP & -0.118852 & 0.0010 \\
DreamSim & AF-FLIP & -0.066221 & 0.2008 \\
CLIP     & AF-FLIP & -0.079379 & 0.0736 \\
FLIP     & AF-FLIP & -0.079379 & 0.0736 \\
FaceNet  & AF-FLIP & -0.171484 & $<.0001$ \\
CosFace  & AF-FLIP & -0.145168 & $<.0001$ \\
ArcFace  & AF-FLIP & -0.118852 & 0.0010 \\
AF-CLIP  & AF-FLIP & -0.032787 & 0.9382 \\
\hline
\end{tabular}
\end{table}

\begin{table}[!h]
\centering
\small
\caption{Attribute importance and $R^2$ of GAM models fitted to Own-group and Other-group human judgments.}
\label{tab:gam_importance}
\setlength{\tabcolsep}{6pt} 
\renewcommand{\arraystretch}{0.9} 
\begin{tabular}{rcc}
\toprule
\textbf{Attribute} & \textbf{Own-group} & \textbf{Other-group} \\
\midrule
Eyebrow shape      & 0.035 & 0.033 \\
Eyebrow thickness  & 0.027 & 0.029 \\
Eye shape          & 0.043 & 0.079 \\
Eye size           & 0.097 & 0.028 \\
Eye color          & 0.076 & 0.056 \\
Nose shape         & 0.020 & 0.063 \\
Nose size          & 0.014 & 0.033 \\
Mouth size         & 0.016 & 0.025 \\
Lip thickness      & 0.016 & 0.026 \\
Lip color          & 0.033 & 0.032 \\
Hair color         & 0.041 & 0.030 \\
Skin color         & 0.015 & 0.070 \\
Skin texture       & 0.016 & 0.081 \\
\midrule
Face aspect ratio   & 0.159 & 0.098 \\
Hair length         & 0.203 & 0.071 \\
Forehead height     & 0.109 & 0.083 \\
Pupillary distance  & 0.020 & 0.054 \\
Cheek shape         & 0.029 & 0.042 \\
Jaw width           & 0.016 & 0.033 \\
Chin shape          & 0.015 & 0.034 \\
\midrule
\textbf{Overall GAM $R^2$} & \textbf{0.520} & \textbf{0.480} \\
\bottomrule
\end{tabular}
\end{table}
% \begin{table}[!h]
% \centering
% \caption{Tukey HSD post-hoc comparison (LSMeans, $\alpha = 0.05$). Methods not sharing a letter are significantly different.}
% \label{tab:tukey_hsd}
% \begin{tabular}{rccc}
% \toprule
% Method & LSMean & Std. Error & Groups \\
% \midrule
% AF-FLIP   & 0.86885 & 0.02257 & A \\
% AF-CLIP   & 0.83607 & 0.02257 & A B \\
% DreamSim  & 0.80263 & 0.02022 & A B C \\
% FLIP      & 0.78947 & 0.02022 & A B C D \\
% CLIP      & 0.78947 & 0.02022 & A B C D \\
% LPIPS     & 0.75000 & 0.02022 & B C D E \\
% ArcFace   & 0.75000 & 0.02022 & B C D E \\
% CosFace   & 0.72368 & 0.02022 & C D E \\
% MoCo      & 0.72368 & 0.02022 & C D E \\
% DINO      & 0.72368 & 0.02022 & C D E \\
% FaceNet   & 0.69737 & 0.02022 & D E \\
% SSIM      & 0.69737 & 0.02022 & D E \\
% PSNR      & 0.68421 & 0.02022 & E \\
% SimCLR    & 0.65789 & 0.02022 & E \\
% \bottomrule
% \end{tabular}
% \end{table}

% \begin{table}[!h]
% \centering
% \vspace{-0.2cm}
% \caption{
% \toreview{Pearson Correlation $\rho$ between inter-landmark distance and model-predicted configural attribute distance by attribute (4 shown).}
% }
% \vspace{-0.3cm}
% \small
% \setlength{\tabcolsep}{5pt}
% \renewcommand{\arraystretch}{0.5}
% \begin{tabular}{cccc}
% \toprule
% Eyebrow thickness & Nose size & Pupillary distance & Jaw width \\
% \midrule
% 0.916 & 0.969 & 0.892 & 0.947 \\
% \bottomrule
% \end{tabular}
% \label{tab:manipulation_check}
% \vspace{-0.3cm}
% \end{table}
\clearpage

\subsection{Module Justification and Validation}
% ---------- Rebuttal (e31B), exact wording from Table R.1 ----------:
% "See Table~R.\ref{tab:design_rationale}."
\rebuttalpoint{Justify specific designs of each module (\e31B).}
\rev{We justify the design rationale of each module in Section~\ref{sec:alignface_model} against standard alternatives in Appendix Table~\ref{tab:design_rationale}.}

\begin{table}[!h]
\centering
\small
\vspace{-0.15cm}
\caption{Justifications for AlignFace modules compared to standard alternatives.}
\vspace{-0.2cm}
\begin{tabular}{p{3.2cm} p{2.7cm} p{7.3cm}}
% \begin{tabular}{>{\raggedright\arraybackslash}p{1.6cm} >{\raggedright\arraybackslash}p{1.6cm} >{\raggedright\arraybackslash}p{4.8cm}}
\hline
\textbf{Module} & \textbf{Alternative} & \textbf{Justification w.r.t. alternative (Benefit)} \\ \hline
Vision-Language Model & Feature Engineering & 
% \textit{Open-ended attributes:} scalable and automatic attribute-conditioned comparisons via prompts. 
More scalable: support open-domain (label-free) attributes instead of hard-coded features.
\\ \hline
Attr-Gated Cross-Attention & Feature Concatenation & 
% \todo{\textit{Attribute-specific comparison:} cross-attention operates analogously to a dictionary lookup, while the gate controls attribute contribution.}
Interpretability and dimensionality reduction: constrain reasoning via attention to focus on information most relevant to each attribute, rather than finding relationships across the full face pair without differencing or focusing.
\\ \hline
Concept Bottleneck Model & MLP & 
% \textit{Attribute-based valuation:} enforce reasoning over human-understandable attributes. 
Interpretability: regularizes reasoning with domain-specific attributes, avoiding spurious features.
\\ \hline
Neural GAM & MLP & 
% \textit{Nonlinear stimulus:} capture both nonlinearity and approximate additivity in human face perception. 
Interpretability: regularizes reasoning independently and nonlinearly for each attribute, avoiding spurious relations.
\\ \hline
\end{tabular}
% \vspace{-0.3cm}
\label{tab:design_rationale}
\end{table}

% \subsection{Attribute Manipulation Validation}
% \label{sec:attribute_manipulation}
% ---------- Rebuttal (e31B), exact wording from Table R.1 ----------:
% "We validated that general-purpose VLMs encode facial configural attribute information by comparing AlignFace's predicted attribute distances against ground-truth heuristic inter-landmark distances~\cite{lugaresi2019mediapipe}. Table~R.\ref{tab:manipulation_check} shows high correlation performance."
\rebuttalpoint{Experiment to verify attribute detection (wsLd, jvsS).}
\rev{We validated that general-purpose VLMs encode featural and configural attribute information by comparing AlignFace's predicted attribute distances against ground-truth heuristic inter-landmark distances~\cite{lugaresi2019mediapipe}.
% and reporting Pearson Correlation $\rho$ 
Appendix Table~\ref{tab:manipulation_check} shows high Pearson correlations $\rho$, indicating that AlignFace reliably tracks the intended physical attribute changes.}
% To verify that AlignFace captures the intended facial attributes rather than spurious visual changes, we evaluated model-predicted attribute distances on single-attribute manipulations.
% For attributes with measurable geometric changes, we computed Pearson correlation between the model-predicted attribute distance and the corresponding landmark-based measurement.
% The consistently high correlations indicate that AlignFace tracks the intended physical attribute changes.}
\begin{table}[!h]
\centering
\caption{
Pearson correlation $\rho$ between heuristic physical measurements and model-predicted attribute distances on single-attribute manipulations.
}
\label{tab:manipulation_check}
\small
\setlength{\tabcolsep}{5pt}
\renewcommand{\arraystretch}{0.9}
\vspace{-0.2cm}

\begin{tabular}{
    M{1.4cm}
    M{2.5cm}
    M{1.8cm}
    % >{\centering\arraybackslash}m{1.8cm}
}
\toprule
\textbf{Category} & \textbf{Attribute} & \textbf{Pearson correlation $\rho$} \\
\midrule

\multirow{5}{*}{Featural}
& Eyebrow thickness & 0.916 \\
& Eye size          & 0.970 \\
& Nose size         & 0.969 \\
& Mouth size        & 0.840 \\
& Lip thickness     & 0.989 \\

\midrule

\multirow{4}{*}{Configural}
& Face aspect ratio & 0.934 \\
& Forehead height & 0.963 \\
& Pupillary distance & 0.892 \\
& Jaw width         & 0.947 \\

\bottomrule
\end{tabular}
\end{table}

\subsection{Baseline Implementation Details}
% ---------- Rebuttal (JeJP), exact wording ----------:
% Baseline evaluation details (JeJP). 
% Reply: "We implemented all models in PyTorch (except PSNR/SSIM) using official pretrained checkpoints. Test setup: for each instance in the test set, we randomly group 3 faces to obtain two similarity/distance ratings for each comparison model, emulating the human 2AFC labeling. More details in Table A.7."
% \rebuttalpoint{Baseline evaluation details (JeJP).}
\rev{We describe additional details of the baseline competitors used to evaluate AlignFace in Section~\ref{sec:baselines}. 
All models are implemented in PyTorch, except for PSNR/SSIM, which use scikit-learn. 
Images are scaled to 224 $\times$ 224 by default, except for face recognition models (112 $\times$ 112).
We loaded the official pretrained checkpoint for each backbone during evaluation. 
Faces are preprocessed to fit each model's input size and normalized according to its backbone. 
Feature embeddings are extracted and used to compute cosine similarity, which then determines the 2AFC triplet response. 
Appendix Table~\ref{tab:baseline_details} shows the backbone and pretrained dataset for each baseline.}
% Appendix Table~\ref{tab:baseline_comparison} presents a conceptual and architectural comparison between AlignFace and these baseline methods.
% }
% \end{multicols}
\begin{table}[!h]
    \centering
    \caption{Baseline implementation details.}
    \label{tab:baseline_details}
    \small
    \setlength{\tabcolsep}{6pt} 
    \renewcommand{\arraystretch}{0.95} 
    \begin{tabular}{r c c}
        \toprule
        \textbf{Baseline} & \textbf{Backbone} & \textbf{Pretrained} \\
        \midrule
        PSNR/SSIM   & ---         & --- \\
        \midrule
        SimCLR      & ResNet-50   & ImageNet-1K \\
        MoCo        & ResNet-50   & ImageNet-1K \\
        DINOv2      & ViT-B/32    & LVD-142M \\
        CLIP        & ViT-B/32    & WIT-400M \\
        FLIP        & ViT-B/32    & FLIP-80M \\
        LPIPS       & AlexNet     & ImageNet-1K \\
        DreamSim    & Ensemble    & ImageNet + synthetic \\
        \midrule
        FaceNet     & Inception-ResNet-v1     & MS1MV3 \\
        CosFace     & Inception-ResNet-v1     & MS1MV3 \\
        ArcFace     & Inception-ResNet-v1     & MS1MV3 \\
        \bottomrule
    \end{tabular}
\end{table}
\newpage

\subsection{Supplementary Results}
\begin{figure}[!ht]
    \centering
    \vspace{-0.2cm}
    \includegraphics[width=\linewidth]{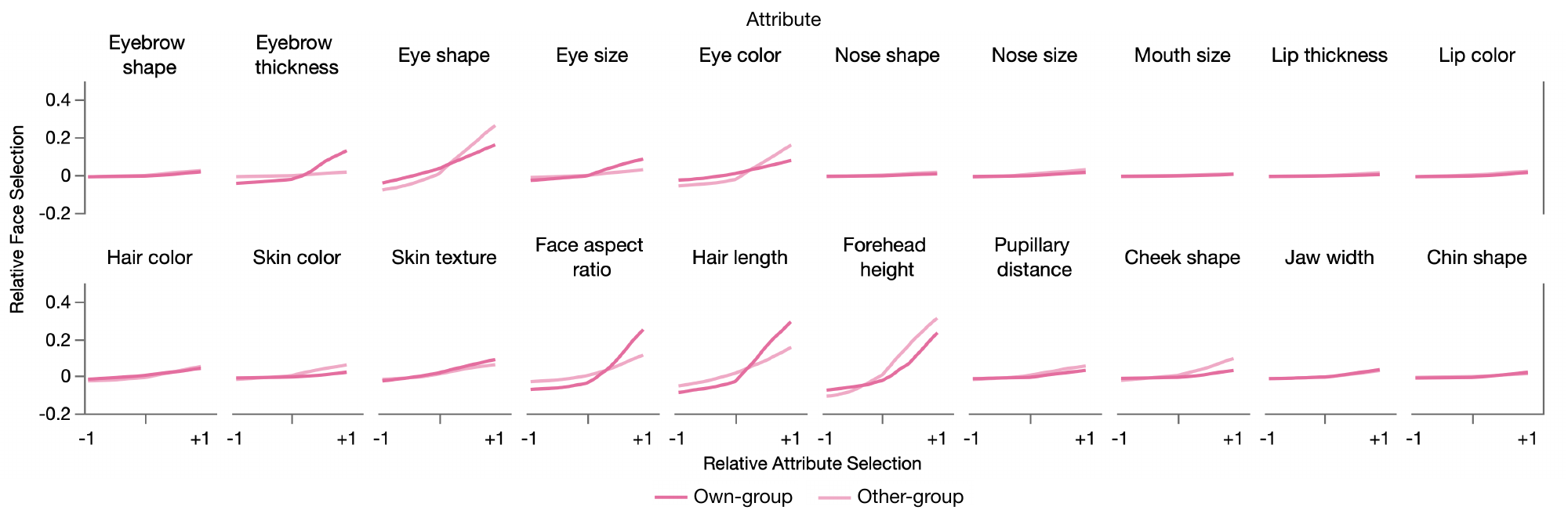}
    \vspace{-0.5cm}
    \caption{
    Partial dependence plots of overall relative face
    selection by relative attribute selection across 20 attributes.
    }
    \vspace{-0.0cm}
    \label{s_fig:gam_alignment_full}
\end{figure}

\begin{figure}[!ht]
    \centering
    % \vspace{-0.2cm}
    \includegraphics[width=\linewidth]{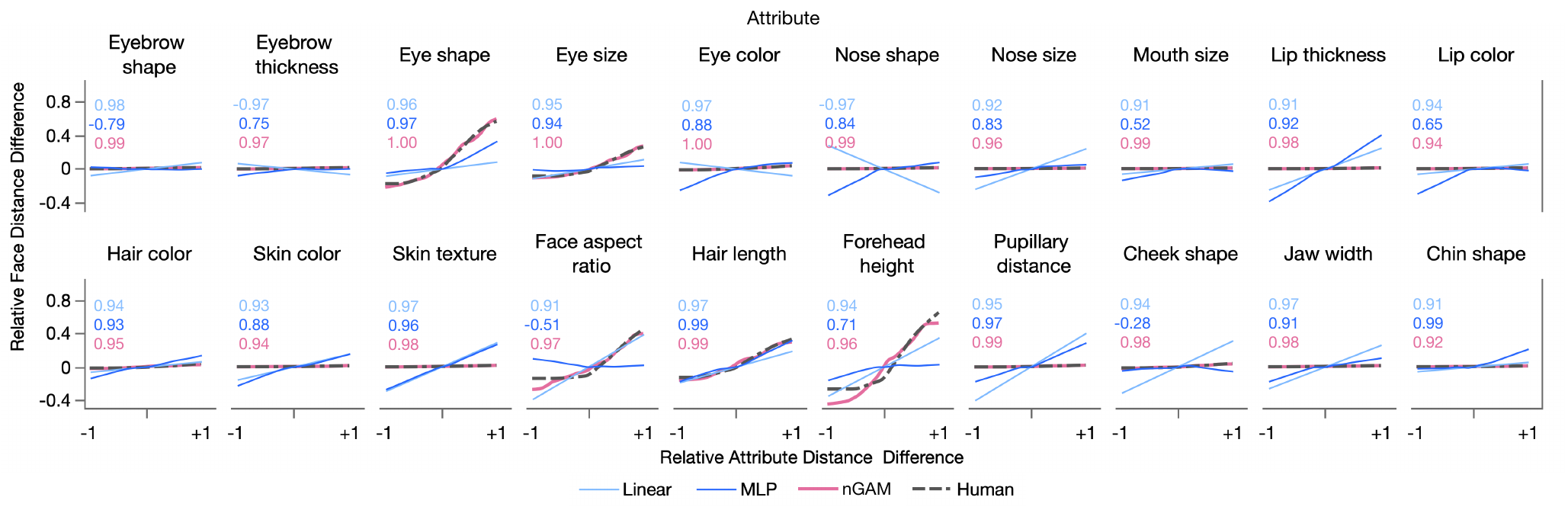}
    \vspace{-0.3cm}
    \caption{Partial dependence plots (PDPs) of overall relative face distance difference by relative attribute distance difference across 20 attributes.
    Numbers denote the Pearson correlation between human judgments and comparators.
    }
    \vspace{-0.5cm}
    \label{s_fig:gam_alignment}
\end{figure}
\clearpage

\begin{figure}[!ht]
    \centering
    \includegraphics[width=0.9\linewidth]{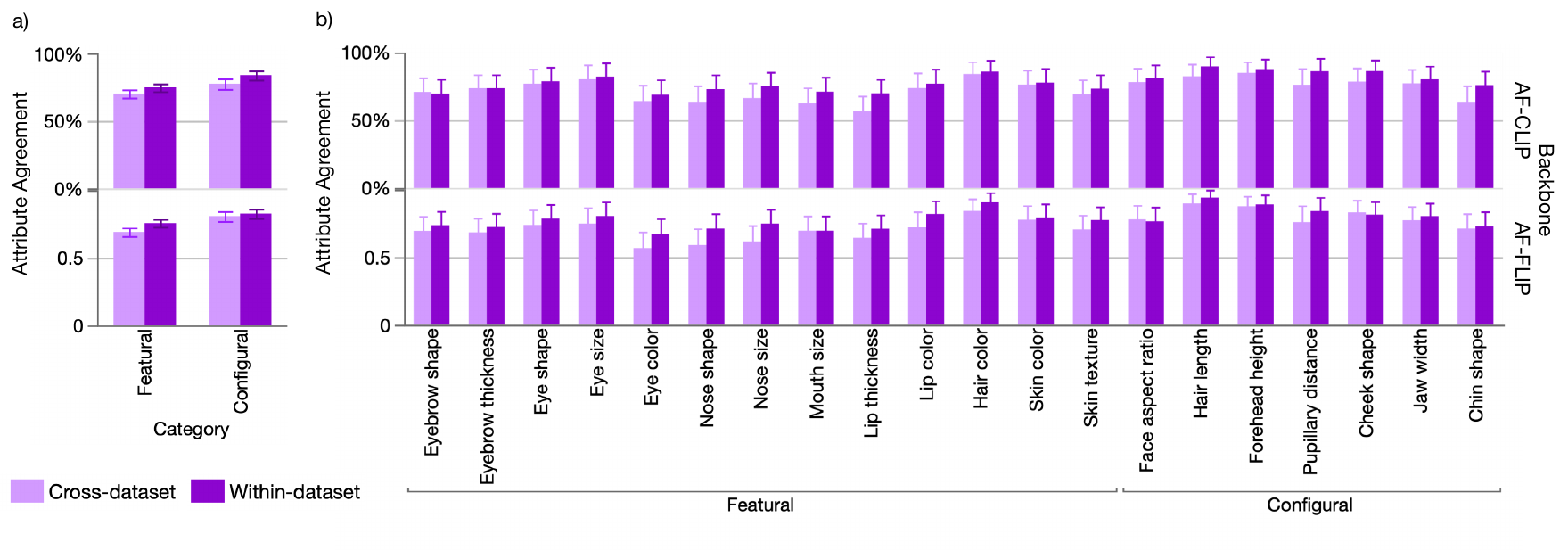}
    \vspace{-0.3cm}
    \caption{Cross-dataset perception agreement for
    a) attribute categories and b) individual attributes.
    }
    \vspace{-0.3cm}
    \label{s_fig:cross_dataset_attr}
\end{figure}

\begin{figure}[!ht]
    \centering
    \includegraphics[width=0.9\linewidth]{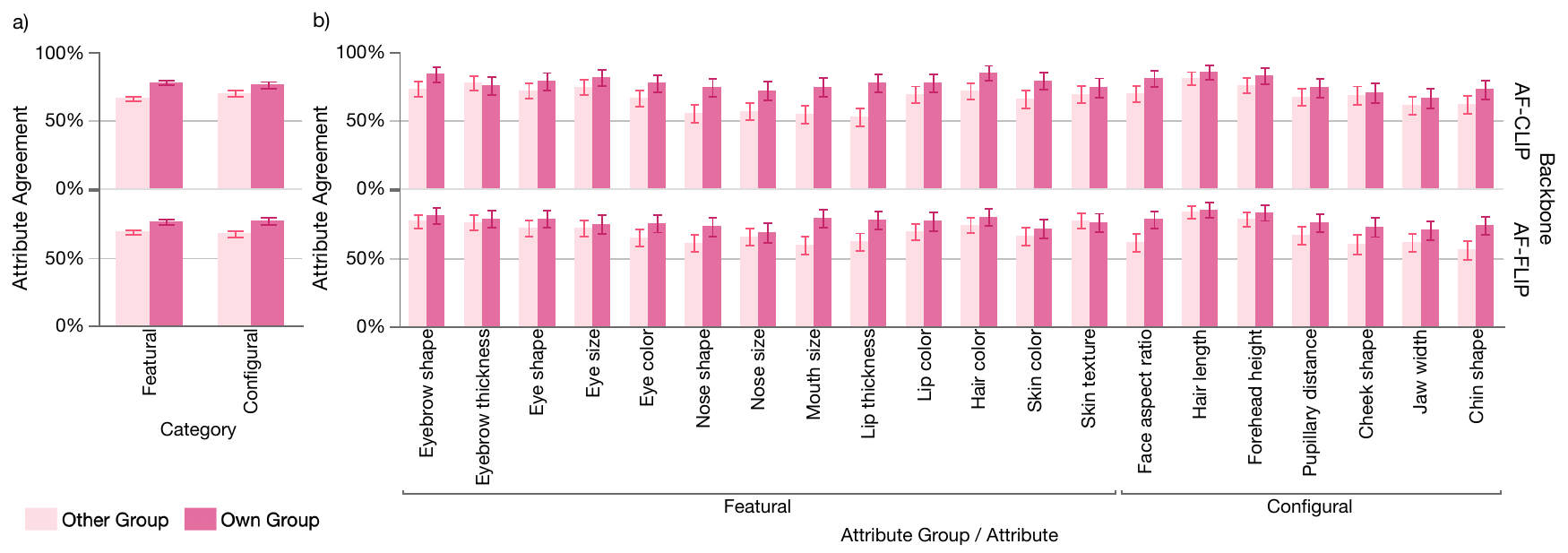}
    \vspace{-0.3cm}
    \caption{Cross-group attribute perception agreement for
    a) attribute categories and b) individual attributes.
    }
    \vspace{-0.3cm}
    \label{s_fig:cross_group_attr}
\end{figure}

\onecolumn

\end{document}